\documentclass[structabstract]{aa}

\usepackage{natbib}
\usepackage{txfonts}

\usepackage{graphicx}
\usepackage{multirow}

\usepackage{hyperref}
\hypersetup{
    final=true,
    pageanchor=true,
    colorlinks=true,
    breaklinks=true,
    linkcolor=blue,
    citecolor=blue,
    urlcolor=blue,
    pdfpagemode=UseNone,
    pdftitle={Horizontal flow fields observed in Hinode G-band images.
        I. Methods},
    pdfauthor={M. Verma and C. Denker},
    pdfsubject={Solar Physics},
    pdfkeywords={Sun: photosphere, Sun: surface magnetism, Sun: sunspots,
        Sun: granulation, Techniques: image processing, Methods: data
        analysis}}

\newcommand\tsp{\mbox{$\;\!$}}
\newcommand\phn{\phantom{0}}

\begin{document}

%###############################################################################
%#
%#    TITLE
%#
%###############################################################################

\title{Horizontal flow fields observed in\\
       \textit{Hinode} G-band images}

\subtitle{I. Methods}

\author{M.\ Verma \and C.\ Denker}

\institute{Astrophysikalisches Institut Potsdam,
    An der Sternwarte 16,
    D-14482 Potsdam,
    Germany\\
    e-mail: \href{mailto:mverma@aip.de}{\texttt{mverma@aip.de}} and 
        \href{mailto:cdenker@aip.de}{\texttt{cdenker@aip.de}}}
%    \email{mverma@aip.de, cdenker@aip.de}}

\date{Received December 20, 2010; accepted March 10, 2011}

\abstract
% context heading (optional)
{The interaction of plasma motions and magnetic fields is an
important mechanism, which drives solar activity in all its facets. For example,
photospheric flows are responsible for the advection of magnetic flux, the
redistribution of flux during the decay of sunspots, and the built-up of
magnetic shear in flaring active regions.}
% aims heading (mandatory)
{Systematic studies based on G-band data from the Japanese
\textit{Hinode} mission provide the means to gather statistical properties of
horizontal flow fields. This facilitates comparative studies of solar features,
e.g., G-band bright points, magnetic knots, pores, and sunspots at various
stages of evolution and in distinct magnetic environments, thus, enhancing our
understanding of the dynamic Sun.}
% methods heading (mandatory)
{We adapted Local Correlation Tracking (LCT) to measure horizontal flow fields
based on G-band images obtained with the Solar Optical Telescope on board
\textit{Hinode}. In total about 200 time-series with a duration between 1--16~h
and a cadence between 15--90~s were analyzed. Selecting both a high-cadence
($\Delta t =15$~s) and a long-duration ($\Delta T=16$~h) time-series enabled us
to optimize and validate the LCT input parameters, hence, ensuring a robust,
reliable, uniform, and accurate processing of a huge data volume.}
% results heading (mandatory)
{The LCT algorithm produces best results for G-band images having a cadence of
60--90~s. If the cadence is lower, the velocity of slowly moving features will
not be reliably detected. If the cadence is higher, the scene on the Sun will
have evolved too much to bear any resemblance with the earlier situation.
Consequently, in both instances horizontal proper motions are underestimated.
The most reliable and yet detailed flow maps are produced using a Gaussian
kernel with a size of 2560~km $\times$ 2560~km and a full-width-at-half-maximum
(FWHM) of 1200~km (corresponding to the size of a typical granule) as sampling
window.}
% conclusions heading (optional)
{Horizontal flow maps and graphics for visualizing the properties of
photospheric flow fields are typical examples for value-added data products,
which can be extracted from solar databases. The results of this study will be
made available within the `small projects' section of the German Astrophysical
Virtual Observatory (GAVO).}

\keywords{Sun: photosphere --
    Sun: surface magnetism --
    Sun: sunspots --
    Sun: granulation --
    Techniques: image processing --
    Methods: data analysis}

\maketitle

%###############################################################################
%#
%#    INTRODUCTION
%#
%###############################################################################

\section{Introduction\label{SEC01}}

Data from space do not suffer the deleterious effects of Earth's turbulent
atmosphere, which blur and distort images so that features might fade into
obscuration, making it difficult to follow them from image to image. The huge
volume of \textit{Hinode} G-band images with good spatial resolution, cadence,
and coverage provide time-series of consistent quality to quantify photospheric
proper motions, which can be used in comparative studies.

Various techniques have been developed in the past decades to
measure horizontal proper motions on the solar surface. Basically, they can be
divided in two classes. The first class includes Feature Tracking (FT) methods,
which follow the footprints of individual features in images of a time-series
\citep[see e.g.][]{Strous1995}. Tracking facular points of opposite magnetic
polarity in an emerging flux region effectively demonstrated the potential of FT
techniques \citep{Strous1996}. However, the image processing steps
(segmentation, labeling, and identification) rely on prior knowledge about the
object under investigation. Therefore, FT methods seem to be better suited for
case studies rather than bulk processing of huge data volumes, where the
reduction of dimensionality is a desirable feature. The balltracking method
developed by \citet{Potts2004} can also be subsumed under FT techniques, since
artificial tracer particles are introduced to follow the footprints of local
intensity minima. The ball tracking method works well for granulation so that it
is a good choice for the characterization of supergranulation \citep{Potts2008}.
Nevertheless, it might introduce a scale dependence, when tracking on other
(small-scale) features such as bright points, penumbral grains, and umbral
dots.

% Figure 1
\begin{figure*}[t]
\centerline{
    \includegraphics[width=0.5\textwidth]{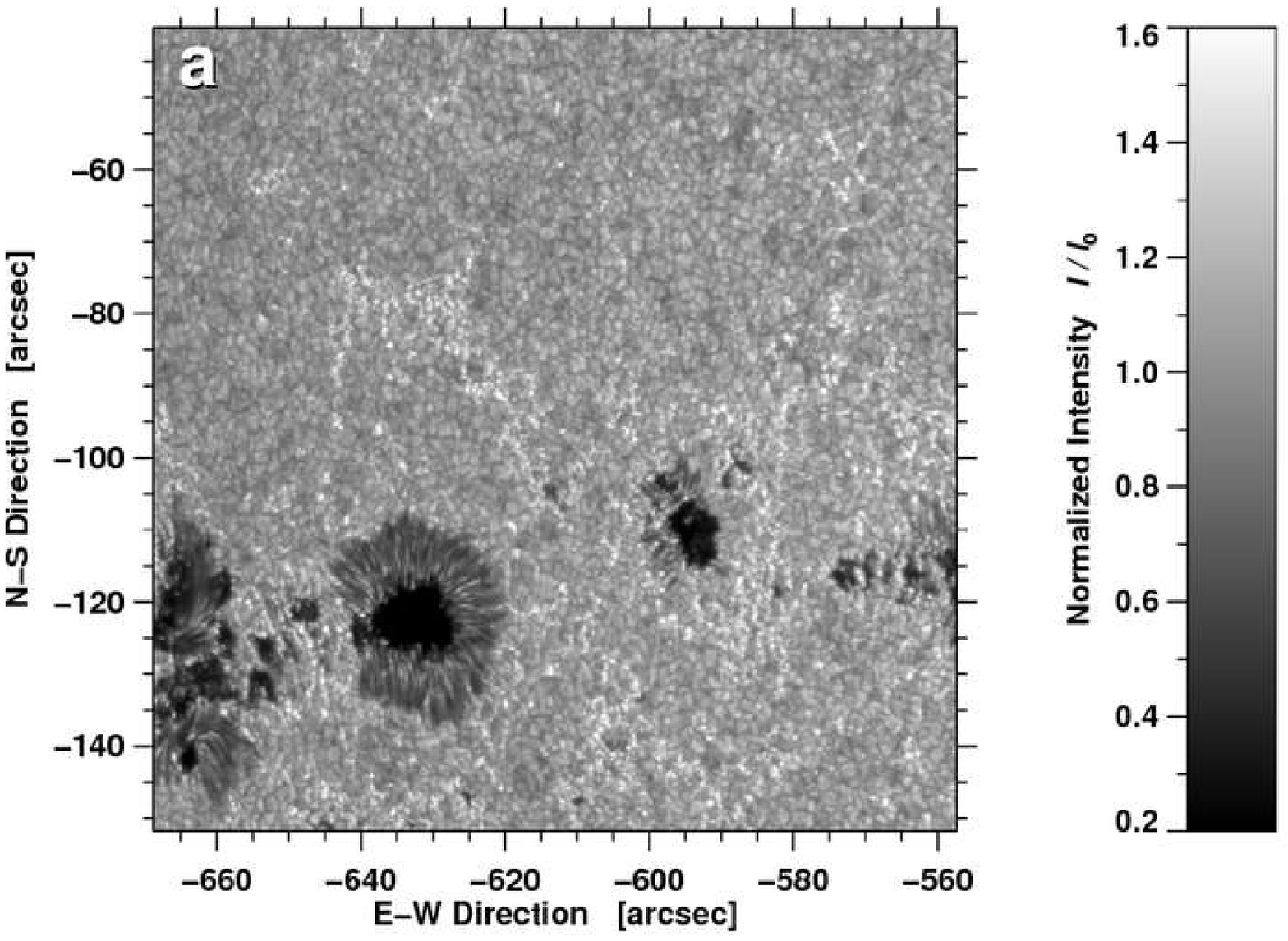}
    \includegraphics[width=0.5\textwidth]{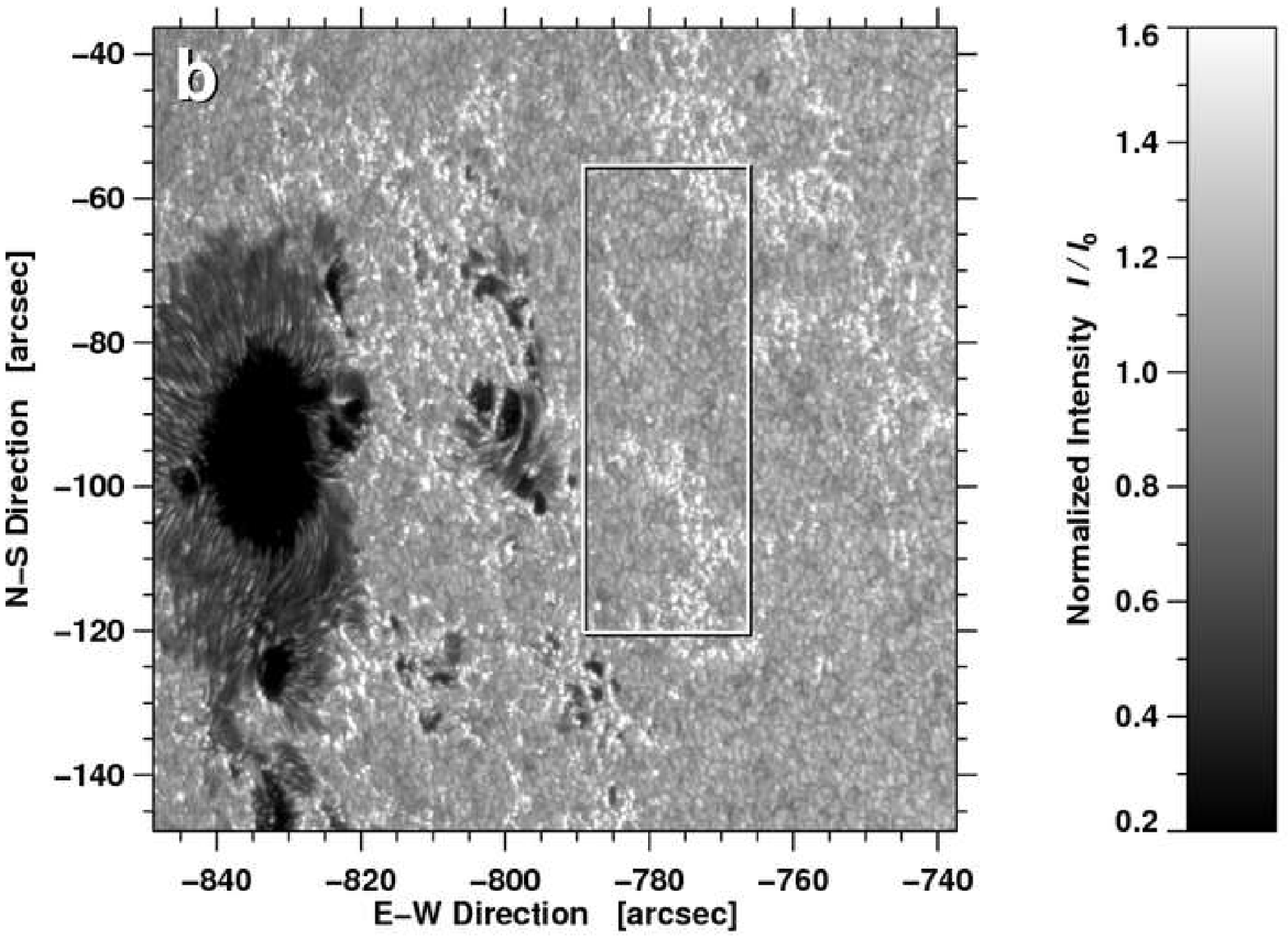}}
\caption{Calibrated G-band images of active regions (a) NOAA~10960 and (b)
    NOAA~10930 observed on 2007 June~4 and 2006 December~7, respectively. The
    FOV is $111\arcsec \times 111\arcsec$. The annotation of the axes refers to
    heliocentric coordinates given in seconds of arc. The region in the
    rectangular white box is used for further analysis zooming in on a quiet Sun 
    region containing numerous G-band bright points outlining supergranular 
    boundaries. The images were normalized such that
    the mode of the quiet Sun intensity distribution corresponds to unity.}
\label{FIG01}
\end{figure*}

The second class is based on Local Correlation Tracking (LCT),
where displacement vectors are derived by cross-correlating small regions in
consecutive images of a time-series. The principles of LCT are laid out in the
seminal work of \citet{November1988}. Leaving behind the underlying velocity
assumption of LCT \citet{Schuck2006} developed a technique to track optical
flows named the differential affine velocity estimators (DAVE). \citet{Chae2008}
presented a formulation of the non-linear case and called it accordingly the
nonlinear affine velocity estimators (NAVE). These authors also provide a
detailed parameter study of LCT, DAVE, and NAVE based on images reflecting
analytical solutions of the continuity equation as well as on magnetogram data
from \textit{Hinode} and MHD simulations. Similarly, \citet{Welsch2007} compared
these velocity inversion techniques based on MHD simulations, where the
velocity field is known.

Various other technical aspects have to be considered while
implementing techniques for optical flow tracking. Since interpolations are
required in the various data reduction steps, we refer to \citet{Potts2003}, who
describe the systematic errors, which can be introduced by unsuitable
interpolation schemes. More recently, \citet{Loefdahl2010} discussed image-shift
measurements in the context of solar wavefront sensors, which are also
applicable to the LCT implementations.

\textit{Hinode} G-band images offer an excellent opportunity
for systematic statistical studies of flow fields, because of their uniform
quality in the absence of seeing distortions, thus alowing us to directly
compare flow fields in different solar settings. In Sect.~\ref{SEC2}, we
introduce the high-cadence and long-duration data sets, which are used to find
the optimal parameters for computing LCT maps based on time-series of
\textit{Hinode} G-band images. The implementation of the algorithm is described
in Sect.~\ref{SEC3}. Section~\ref{SEC4} presents the results of this parameter
study justifying our choice of LCT parameters, which will be used to create
value-added data to complement the \textit{Hinode} database. The most important
parameters are summarized in Sect.~\ref{SEC5}, which also introduces some of the
future uses of such a database.

%###############################################################################
%#
%#    OBSERVATIONS
%#
%###############################################################################

\section{Observations\label{SEC2}}

Images obtained in the Fraunhofer G-band (bandhead of the CH molecule at
$\lambda$430.5~nm) have high contrasts, and small-scale magnetic features can be
simply identified with bright points \citep{Berger1995}. Despite the
observational advantages of such ``proxy-magnetometry'' \citep{Leenaarts2006},
the theoretical description of the molecular line formation process is far from
easy \citep[cf.][]{SanchezAlmeida2001, Steiner2001, Schuessler2003}. LCT
techniques, however, can take full advantage of the high contrast and the rich
structural contents of G-band images. On \textit{Hinode}
\citep{Kosugi2007} such observations are carried out by the Broad-band Filter
Imager (BFI) of the Solar Optical Telescope \citep[SOT,][]{Tsuneta2008}.

Our initial selection criteria were that at least 100 G-band images had to be
recorded on a given day, which should have additionally a cadence of better than
100~s. It turned out that these criteria restricted us to data with only half
the spatial resolution (0.11\arcsec\ pixel$^{-1}$), where $2 \times 2$ pixels
were binned into one. In total 48 data sets with $2048 \times 1024$ pixels and
153 data sets with $1024 \times 1024$ pixels were selected for further analysis.
The time intervals covered by these data sets range from one to 16 hours. The
bulk statistical analysis will be presented in forthcoming publications. Here,
we will discuss in detail the LCT algorithm and justify our choice of input
parameters. For this purpose, we selected two data sets: one with a high cadence
and another one with a long duration. In addition, we picked a data set without
binning to study the dependence of flow maps on the spatial resolution of the
input data.

\subsection{High-cadence sequence\label{SEC2.1}}

LCT depends on several input parameters such as the time interval between
successive images and the sampling window's size and form. We analyzed an
one-hour time-series with a cadence of 15~s to validate the intrinsic accuracy
of the LCT algorithm. The data were captured from 14:27--15:27~UT on 2007 June~4
(see Fig.~\ref{FIG01}a). The time-series contains 238 images with $1024\times
1024$~pixels. The observations were centered on active region NOAA~10960, which
was located on the solar disk at heliocentric coordinates E630\arcsec\ and
S125\arcsec\ ($\mu = 0.75$). The active region was in the maximum growth phase
and had a complex magnetic field configuration. NOAA~10960 was classified as a
$\beta \gamma \delta$-region and was the source of many M-class flares including
a major M8.9 flare at 05:06~UT on 2007 June~4, which has been analyzed in a
multi-wavelength study by \citet{Kumar2010}.

\subsection{Long-duration sequence\label{SEC2.2}}

Solar features evolve on different time scales from about five minutes for
granulation to several tens of hours for supergranulation. Obviously, the time
over which individual LCT maps are averaged plays a decisive role in the
interpretation of such average flow maps. Hence, we selected a time-series with
16 hours of continuous data captured on 2006 December~7 (see Fig.~\ref{FIG01}b).
This long-duration sequence starts at 02:30~UT and ends at 18:30~UT. It includes
960 images with  $1024\times 1024$~pixels and has a cadence of 60~s. Due to
memory constrains imposed by the subsonic filtering we choose an area of $210
\times 595$~pixels centered on a region with granulation and G-band bright
points (white box Fig.~\ref{FIG01}b). This region is to the west of active
region NOAA~10930 located at heliocentric coordinates E777\arcsec and
S88\arcsec\ ($\mu = 0.59$). The sunspot group was classified as a $\beta \gamma
\delta$-region exhibiting a complex magnetic field topology and producing
numerous C-, M-, and X-class flares. This region has been extensively studied,
especially around the time of an X3.4 flare on 2006 December~13
\citep[e.g.,][]{Schrijver2008}. LCT techniques were used by \citet{Tan2009} to
study horizontal proper motions associated with penumbral filaments in a rapidly
rotating $\delta$-spot.

\subsection{High-spatial resolution sequence\label{SEC2.3}}

Only a few data sets of G-band images exist with the full spatial resolution of
0.055\arcsec\ pixel$^{-1}$ and a cadence suitable for LCT. We selected a
one-hour time-series, which was acquired starting at 04:00~UT on 2006
November~26. This high-spatial resolution sequence contains 118 images with
$2048\times 2048$~pixels and has a time cadence of about 30~s. The observations
were centered on a quiet Sun region near disk center at heliocentric coordinates
E108\arcsec\ and S125\arcsec\ ($\mu = 0.99$) containing only a few G-band bright
points and no major magnetic flux concentrations.

%###############################################################################
%#
%#    IMPLEMENTATION OF THE LCT ALGORITHM
%#
%###############################################################################

\section{Implementation of the LCT algorithm\label{SEC3}}

\subsection{Preprocessing of the G-band images\label{SEC3.1}}

%\footnote{\href{http://www.ittvis.com/}{\texttt{www.ittvis.com}}}
%\footnote{\texttt{www.ittvis.com}}

The data analysis is carried out in the Interactive Data Language
(IDL)\footnote{\href{http://www.ittvis.com/}{\texttt{www.ittvis.com}}}. Data sets are split in 60-minute
sequences with 30~min overlap between consecutive sequences. In preparation for
the LCT algorithm basic data calibration is performed, which consists of dark
current subtraction, correction of gain, and removal of spikes caused by high
energy particles. Figure~\ref{FIG01} contains sample G-band images for both the
high-cadence and long-duration data sets after basic calibration. After initial
data calibration, the geometric foreshortening is corrected and images are
resampled in a regular grid with spacing of 80~km $\times$ 80~km, i.e., the
images appear as if observed at the center of the solar disk. Residual effects
of projecting the surface of a sphere onto a plane are neglected, since the FOV
of the G-band images is still relatively small. A grid size of 80~km was chosen
so that the fine structure contents of the G-band images are not diminished.
Pixels close to the solar limb are projected onto several pixels in planar
coordinates. Thus, the accuracy of the flow maps at these locations is not as
good as for locations close to disk center.

In a 60-minute sequence are $l = 0, 1, 2, \ldots, L-1$
calibrated and deprojected images, where $L$ is the total number of images in a
particular sequence. For an image with $N \times M$ pixels the intensity
distribution is represented by $i(x, y)$ with $x = x_0, x_1, \ldots, x_{N-1}$
and $y = y_0, y_1, \ldots, y_{M-1}$ as pixel coordinates. The indices are
typically dropped to ease the notation. The data processing makes extensive use
of the Fast Fourier Transform (FFT). The FFT of the intensity distribution
$i(x,y)$ is simply denoted by ${\cal F}(i(x,y))$.

\subsection{Aligning the images within a time-series\label{SEC3.2}}

In principle, images could be aligned using the pointing information of the
spacecraft. However, we calculate shifts between consecutive images $i_{l-1}(x,
y)$ and $i_l(x, y)$ by computing the cross-correlation using only the central
part of the images, which is half of the original image size.
These shifts are then applied in succession to align all images
with respect to the first image using cubic spline interpolation with subpixel
accuracy. The signature of the 5-minute oscillation is removed from the
time-series by applying a 3D Fourier filter. This filter, sometimes called a
subsonic filter, has a cut-off velocity of $c_s \approx 8$~km~s$^{-1}$
corresponding to the photospheric sound speed. Since the
subsonic filter uses a 3D Fourier transform some edge effects are sometimes
noted for the first and last few images of a time-series. For this reason we
decided to discard the images during the first and last two minutes of the
time-series after applying the subsonic filter. Therefore, the final time-series
is shortened by this amount of time (see Sect.~\ref{SEC3.5}).

\subsection{LCT algorithm\label{SEC3.3}}

The LCT algorithm is based on ideas put forward by \citet{November1988}.
The algorithm has been adapted to subimages with sizes of $32
\times 32$ pixels corresponding to 2560~km $\times$ 2560~km, so that structures
with dimensions smaller than a granule will contribute to the correlation
signal. Since cross-correlation techniques are sensitive to strong intensity
gradients, a high-pass filter was applied to the entire image suppressing
gradients related to structure larger than granules. The high-pass filter is
implemented as a Gaussian with a FWHM of 15 pixels (1200~km). To indicate that
we refer to an subimage with 32 $\times$ 32 pixels and not the entire image, we
use the notation $i(x^\prime, y^\prime)$. The Gaussian kernel used in the
high-pass filter then becomes
\begin{equation}
g(x^\prime, y^\prime) = \frac{1}{2\pi\sigma^2} 
    \exp\left(-\frac{r(x^\prime, y^\prime)^2}{2\sigma^2}\right)
    \quad,
\end{equation}
where $\sigma = \mathrm{FWHM} / (2\sqrt{2\ln2})$ and $r(x^\prime, y^\prime) =
(x^{\prime\;\! 2} + y^{\prime\;\! 2})^{1/2}$.
The high-pass filtered image can be expressed as
\begin{equation}
i_\mathrm{high}(x, y) = i(x, y) - i(x, y) \otimes g(x^\prime, y^\prime) \quad,
\end{equation}
where $\otimes$ denotes a convolution. The result is an image rich in detail,
where the low spatial frequencies have been removed.

The core of the LCT algorithm is the cross-correlation $c_l(x,
y, x^\prime, y^\prime)$ computed over a $32 \times 32$ pixel region centered on
the coordinates $(x, y)$ for each pixel in image pairs $i_{l-1}(x, y)$ and
$i_l(x, y)$, which can be written as
\begin{eqnarray}
c_l(x, y, x^\prime, y^\prime) & = & \Re\big\{
    {\cal F}^{-1}  \big[ {\cal F} \big( i_{l-1}(x, y, x^\prime, y^\prime)
    g(x^\prime, y^\prime) \big) \cdot \nonumber\\
    &   &
    {\cal F}^{\ast} \big( i_l(x, y, x^\prime, y^\prime)
    g(x^\prime, y^\prime) \big) \big] \big\} 
    d(x^\prime, y^\prime) \quad,
\end{eqnarray}
where $g(x^\prime, y^\prime)$ denotes a weighting function also serving as an
apodising window. This function has the same form as the Gaussian kernel
previously used in the high-pass filter. This ensures that the displacement
vectors are computed without preference in azimuthal direction.
We also multiplied the cross-correlation functions by a mask
$d(x^\prime, y^\prime)$ so that the maximum of the cross-correlation function is
forced to be within a distance of $c_{s, {\rm lim}} = 12$~pixels from its
center. The typical time interval between consecutive images is in the range
from 60--90~s, i.e., a feature moving at the photospheric sound speed of $c_s
\approx 8$~km~s$^{-1}$ would travel 480--720~km corresponding to 6--9 pixels.
This justifies our choice of $c_{s, {\rm lim}}$, which also takes into account
some numerical errors. The position of the maximum of the cross-correlation
function is calculated with subpixel accuracy by a parabolafit to the
neighboring pixels. The numerical accuracy of the parabola fit is about one
fifth of a pixel or 16~km on the solar surface, which corresponds to about
200~m~s$^{-1}$ for proper motions measured from a single pair of G-band images.
Therefore, many flow maps have to be averaged to determine reliable horizontal
proper motions.

\subsection{LCT data products\label{SEC3.4}}

\begin{figure*}[t]
\centerline{
    \includegraphics[width=0.45\textwidth]{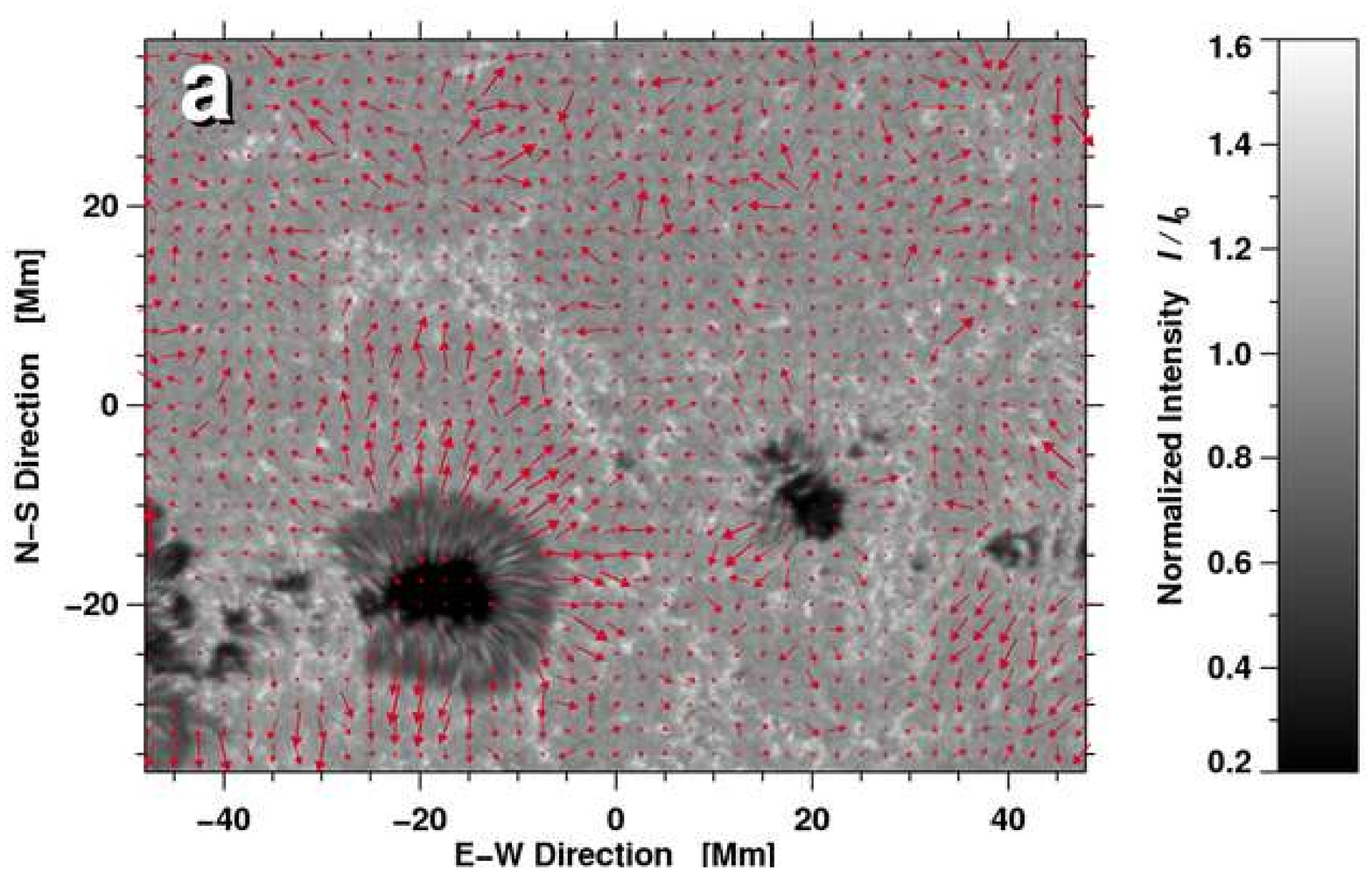}
    \includegraphics[width=0.45\textwidth]{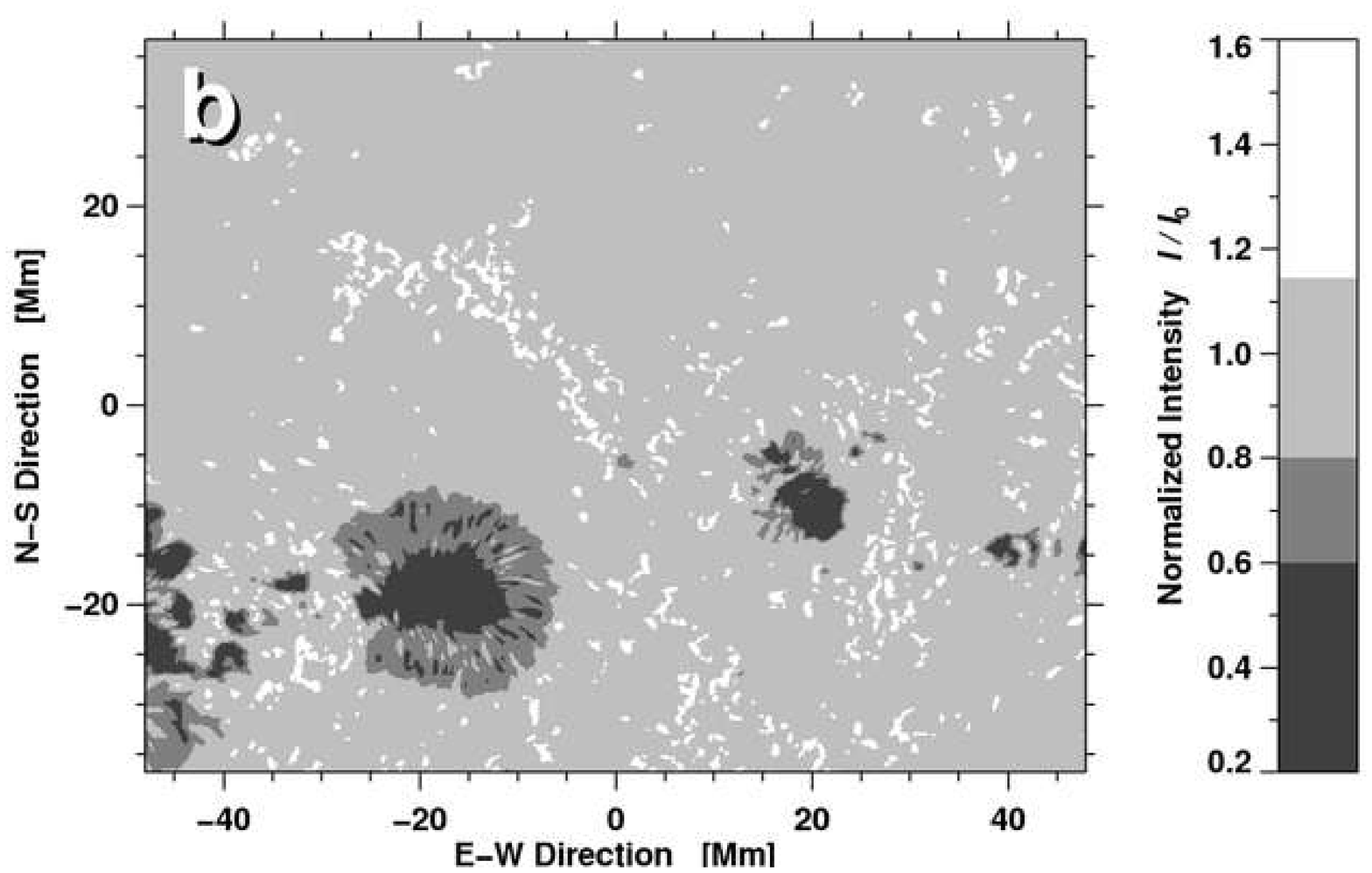}}
\centerline{
    \includegraphics[width=0.45\textwidth]{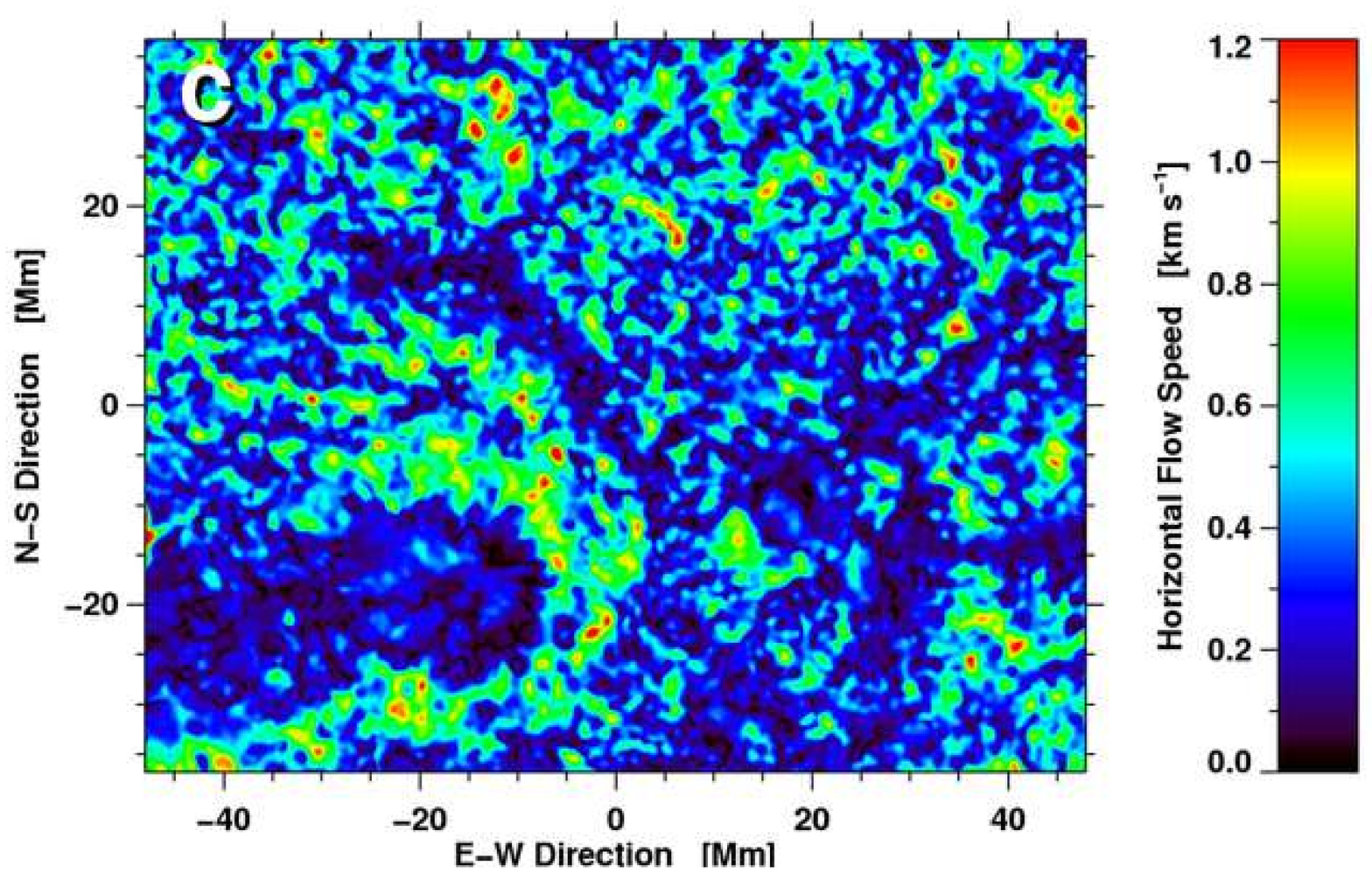}
    \includegraphics[width=0.45\textwidth]{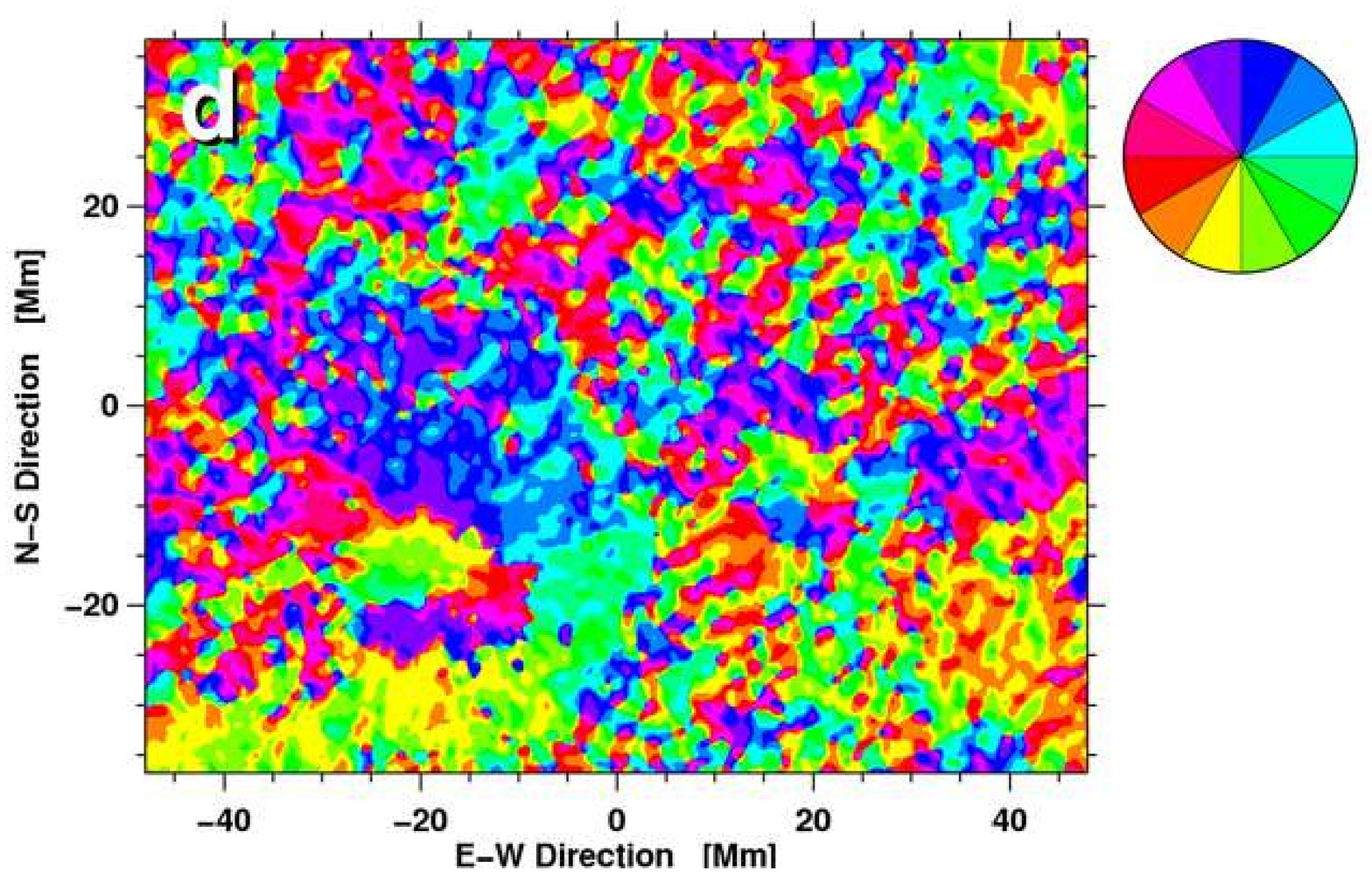}}
\centerline{
    \includegraphics[width=0.45\textwidth]{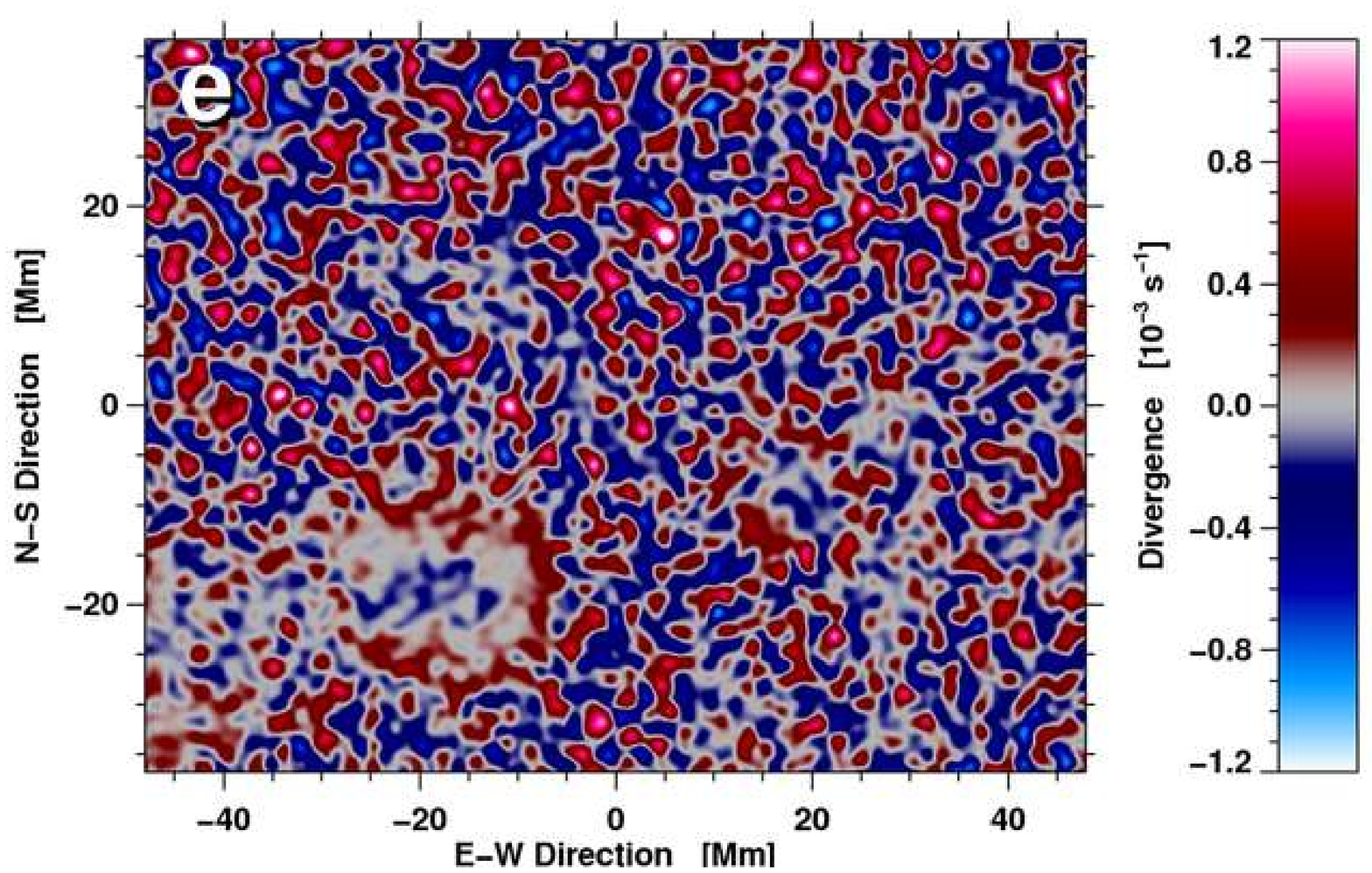}
    \includegraphics[width=0.45\textwidth]{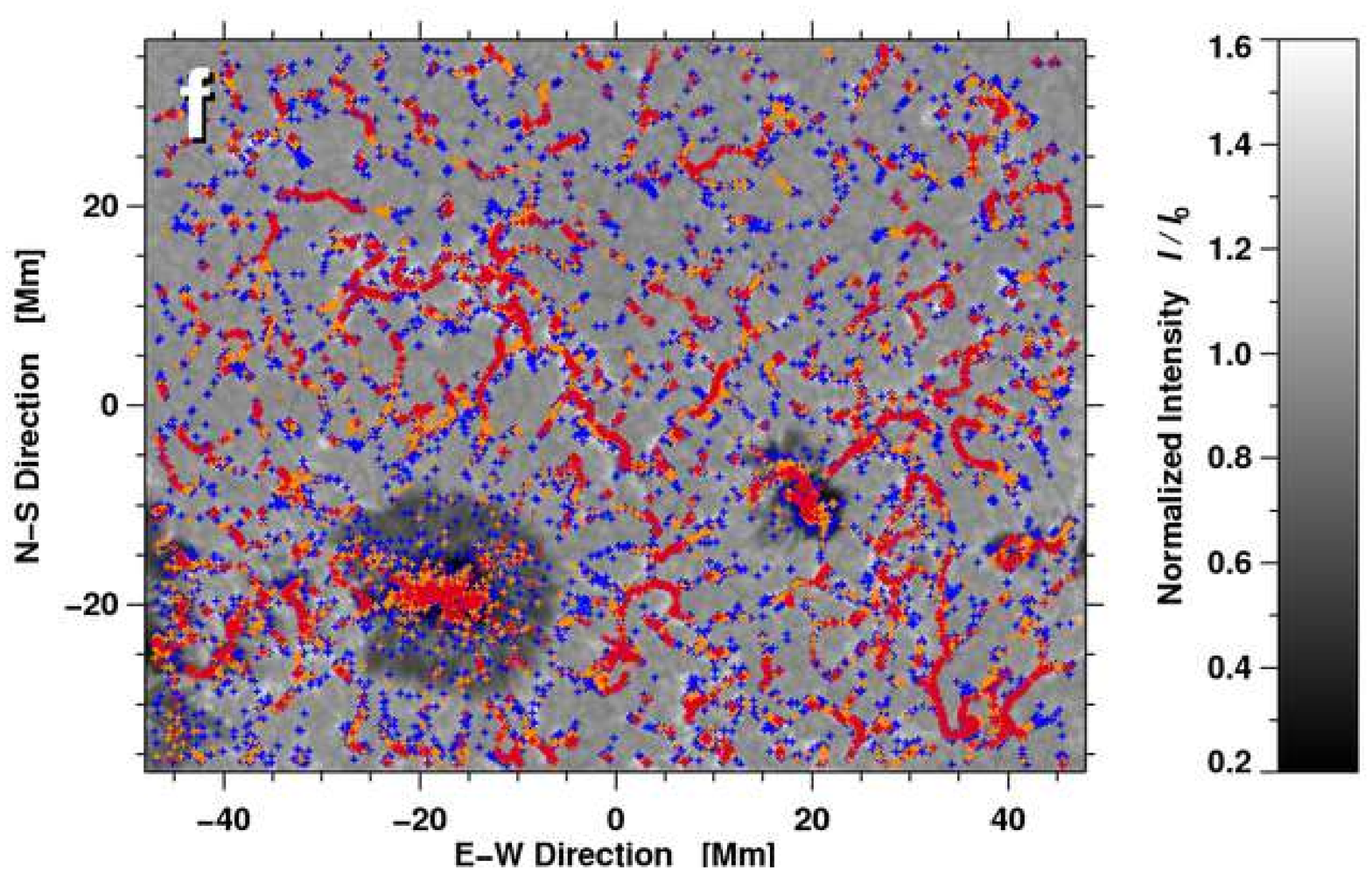}}
\caption{(a) The average (60-minute) G-band image after correction of
    geometrical foreshortening corresponding to the G-band image shown in
    Fig.~\ref{FIG01}a. The red arrows indicate magnitude and direction of the
    horizontal proper motions. Arrows with a length corresponding to the grid
    size indicate velocities of 1~km~s$^{-1}$. (b) Adaptive and fixed intensity
    thresholds are used to identify conglomerates of G-band bright points,
    granulation, penumbrae, umbrae, and pores. Color codes are used to point out
    (c) flow speed and (d) direction in high-resolution flow maps. The flow
    direction is encoded according to the 12 colors of the compass rose. Sources
    and sinks in a flow field can be identified in (e) a divergence map. The (f)
    forward cork maps provides further means of visualizing converging motions.
    Test particles are displayed after they have followed the
    flows for 2 (blue), 4 (orange), and 8~h (red). The conspicuous network of
    corks is related to the spatial scales of the meso-and supergranulation.}
\label{FIG02}
\end{figure*}

Once the individual flow maps are calculated they are saved in binary format. In
addition, average maps of horizontal speed and flow direction as well as the
$x$- and $y$-components of the horizontal flow velocity $(v_x, v_y)$ are stored
in native IDL format. Some auxiliary variables are saved as well so that they
can be used, e.g., in annotating plots depicting the flow fields. A sample of
such plots for a 60-minute average flow field is shown in Fig.~\ref{FIG02}. In
Fig.~\ref{FIG02}a horizontal proper motions are plotted as red arrows with a
60-minute averaged G-band image as a background. The moat flow starting at the
sunspot penumbra and terminating at the surrounding G-band network is clearly
discernible.

In Fig.~\ref{FIG02}b we used an adaptive thresholding algorithm to discern
between granulation, G-band bright points, and strong magnetic features.
Indiscriminately, we used a fixed intensity threshold of $I_\mathrm{mag} = 0.8$
for strong magnetic features and an adaptive threshold for G-band bright points,
which can be given as
\begin{equation}
I_\mathrm{bp} = 1.15 + 0.2(1-\mu) \quad,
\label{EQN04}
\end{equation}
where $\mu = \cos(\theta)$ is the cosine of the heliocentric angle $\theta$. The
darkest parts of sunspots (umbrae) and pores can be identified using another
fixed threshold of $I_\mathrm{dark} = 0.6$, while sunspot penumbrae cover
intermediate intensities from $I_\mathrm{dark}$ to $I_\mathrm{mag}$, allotting
the range $I_\mathrm{mag}$ to $I_\mathrm{bp}$ to granulation. The adaptive
threshold was necessary, as a first order approximation, to account for the
center-to-limb variation (CLV) of the G-band bright points, which exhibit much
higher contrasts near the solar limb. This adaptive thresholding algorithm
allows us to study the properties of horizontal proper motions for different
solar features.

% Figure 2

Apart from the conventional way of displaying velocity vectors as arrows we
present two-dimensional high-resolution speed (Fig.~\ref{FIG02}c) and azimuth
(Fig.~\ref{FIG02}d) maps. In these maps, the physical quantities are computed
for each individual pixel so that the fine structure of the flow field becomes
accessible. The color scale for the speed values is the same for all plots in
this study. Indeed, we used the same color scale for all flow maps in the
database so that flows for different scenes on the Sun can be directly compared.
In the azimuth map the direction is encoded in a 12-color compass-rose, which
can be found to the very right of this panel. In principle more colors could be
used to illustrate the flow direction. However, such plots would become very
crowded and are very hard to interpret. The essential features of the flow
field, i.e., inward motion of the penumbral grains in the inner penumbra and
outward motions related to Evershed and moat flows, can easily be identified.

Divergence (Fig.~\ref{FIG02}e) and vorticity maps were also
compiled for each time-series. The maximum values for divergence ($1.5 \times
10^{-3}$~s$^{-1}$) and vorticity ($0.6 \times 10^{-3}$~s$^{-1}$) in these maps
are an order of magnitude higher than values found for quiet Sun granulation in
previous studies. However, higher values are not surprising, since we calculated
them for each pixel in the high-resolution maps capturing more of the
small-scale motions. Reassuringly, our values of the $10^\mathrm{th}$ percentile
for divergence ($4.3 \times 10^{-4}$~s$^{-1}$) and vorticity ($1.9 \times
10^{-4}$~s$^{-1}$) are essentially the same as reported by \citet{Simon1994}. On
the other hand, in the present study the divergence is four times and the
vorticity is two times higher than the values presented by \citet{Strous1996}
but these values were computed in the proximity of an emerging active region. In
summary and keeping in mind the different LCT parameters used in the different
studies, the statistical properties of the flow fields as presented in this
study are in agreement with previous investigations.

In order to visualize the temporal evolution of the horizontal
proper motions, we computed forward (Fig.~\ref{FIG02}f) and inverse cork maps.
Evenly spread test particles are allowed to float forward in time with a given
horizontal speed for a certain time interval \citep[see][]{Molowny-Horas1994a}.
In the inverse cork map the particles float backward in time. This is
accomplished by simply reversing the sign of the velocity components. The
forward cork map is used to visualize regions of converging flows and the
inverse map is a good tool to study divergence regions. We tracked test
particles for consecutively two, four, and eight hours. These particles were
initially distributed on an equidistantly spaced grid with a spacing of
10~pixels, i.e., one particle was placed every 0.8~Mm. The most conspicuous
feature of the forward cork map are the tracer particles outlining the network
of G-band bright points, which corresponds to the supergranular boundaries.

We prepared overview web pages for the respective dates, when
suitable time-series of G-band images were available, which contain all the six
plots of Fig.~\ref{FIG02} along with vorticity and inverse cork maps. The
results of this study will ultimately be published as a small
GAVO\footnote{\href{http://www.g-vo.org/}{\texttt{www.g-vo.org}}} project as a value-added product of the
\textit{Hinode} database.

%\footnote{\texttt{www.g-vo.org}}
%\footnote{\href{http://www.g-vo.org/}{\texttt{www.g-vo.org}}}

\subsection{Timing issues related to the image capture\label{SEC3.5}}

LCT delivers localized displacements observed in image pairs. These
displacements in conjunction with the time, which has elapsed between the
acquisition of both images, yield localized velocity vectors. Therefore,
accurate knowledge of the time interval between consecutive images used in the
LCT algorithm is essential. The time interval $\Delta t$ for
the high-cadence image sequence observed on 2007 June~4 has a bimodal
distribution with values $ \Delta t$ of 14.4~s (60.7\%) and 16~s (39.3\%). The
average value is $\Delta \bar{t} \approx 15.0$~s. The difference of 1.6~s is an
artifact of the polarization modulation. The polarization modulation unit (PMU)
is located just behind the telescope exit slit within the optical telescope
assembly but before the tip-tilt mirror employed by the correlation tracker
\citep[see][]{Tsuneta2008}. A common CCD camera is assigned to both the broad-
and narrowband filter imagers (BFI and NFI). The critical timing between camera
and PMU is handled by the focal plane package. The PMU is a continuously
rotating waveplate, which is always turned on -- even for non-polarimetric data
such as G-band images. Its rotation period is $T = 1.6$~s and all exposure
timing is controlled with the clock of the PMU. This is the reason for the
non-uniformity in the observed time-interval $\Delta t$.

However, we do not find a bimodal pattern in the LCT
displacements but only fluctuations related to evolving features on the Sun and
some residual numerical effects. Therefore, we opted to use the fixed time
interval $\Delta \bar{t}$ in the data analysis. Sometimes a `traffic jam' in the
data transfer might result in even largertiming errors. On the other hand,
averaging individual LCT maps over an hour (or longer) will only result in
velocity errors of less than a tenth of a precent, i.e., the
speed measured by the LCT algorithm is not significantly affected. In summary,
accurate timing has to be ensured to obtain reliable LCT flow maps and all data
were checked for consistency between recorded time stamps and measured
horizontal displacements. Since the individual flow maps during the first and
last two minutes of the one-hour sequences do not reflect the true proper
motions but are artifacts of the subsonic filtering, we excluded them from the
calculation of the average flow maps.

%###############################################################################
%#
%#    RESULTS
%#
%###############################################################################

\section{Results\label{SEC4}}

\subsection{Statistical properties of flow maps and time cadence
    selection\label{SEC4.1}}

The one-hour time-series on 2007 June~4 contains 238 G-band images, i.e., the
time cadence is $\approx 15$~s. This higher temporal resolution allows us to
study the intrinsic accuracy of the LCT algorithm. We calculated LCT maps using
seven different time intervals $\Delta t=$ 15, 30, 60, 90, 120, 240, and 480~s.
Note that the time interval over which flow maps are averaged is reduced to
$\Delta T = 3600~\mathrm{s} - \Delta t$, i.e., in case of the longest time
cadence by as much as 8~min. However, these slightly different averaging times
will not change the results discussed below. When using all individual LCT maps
to compute the average horizontal proper motion we refer to such data as the
\textit{entire sequence}. On the other hand, when we split the entire sequence
into four disjunct sets, we refer to them as \textit{interleaved data sets},
i.e., every fourth LCT map is employed to compute the average horizontal proper
motion. Since these flow maps cover exactly the same period of time, differences
can be directly attributed to the numerical accuracy of the LCT algorithm.

For the entire sequence and interleaved data sets we computed statistical
parameters describing the distribution of horizontal proper motions for
granulation in the vicinity of active region NOAA~10960. We used the adaptive
thresholding algorithm (Eqn.~\ref{EQN04}) to select only granulation excluding
G-band bright points. The proper motions, thus, refer to plasma motions in the
absence of any strong magnetic field concentrations. Such selection facilitates
comparing horizontal flow speeds and their distributions in all the cases of the
present work.

We calculated the mean $\bar{v}$, median $v_\mathrm{med}$, maximum
$v_\mathrm{max}$, and $10^\mathrm{th}$ percentile $v_{10}$ of the horizontal
flow speeds. Since the maximum speed $v_\mathrm{max}$ is only based on a single
value, it is easily influenced by numerical errors and the data calibration. The
$10^\mathrm{th}$ percentile $v_{10}$ is more robust because it describes a
property of the entire distribution, i.e., the high velocity tail. Along with
these quantities we also calculated the variance $\sigma^2_v$, standard
deviation $\sigma_{v}$, skewness $\gamma_{1,v}$, and kurtosis $\gamma_{2,v}$.
The last two statistical parameters describe the deviation of
the distribution from a normal distribution.

% Figure 3
\begin{figure}[t]
\includegraphics[width=0.5\textwidth]{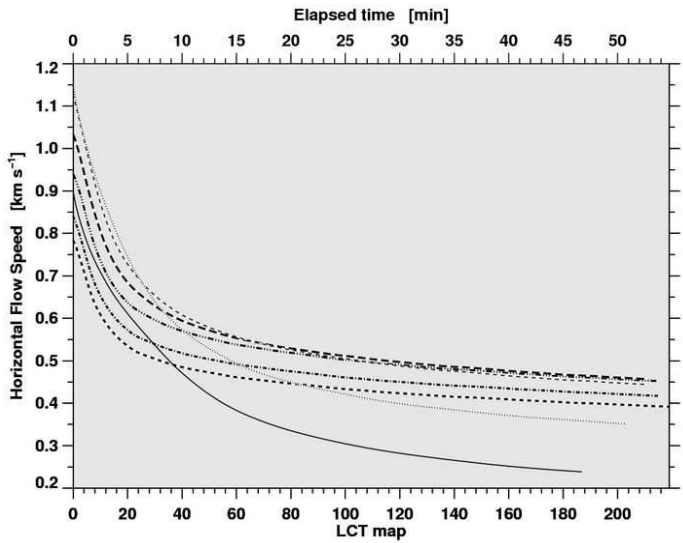}
\caption{Mean horizontal flow speed as a function of averaged LCT maps for time
    cadences $\Delta t$ of 15~s (thick dashed), 30~s (thick dash-dotted curve),
    60~s (thick dash-triple-dotted curve), 90~s (thick long dashed curve), 120~s
    (thin dashed), 240~s (thin dotted), and 480~s (thin solid).}
\label{FIG03}
\end{figure}

We find an increase of the average velocity with increasing time cadence $\Delta
t$ starting from about 0.40~km~s$^{-1}$ for $\Delta t=15$~s, arriving at a
maximum value of about 0.47~km~s$^{-1}$ for $\Delta t=$ 60--90~s, and then
decreasing from about 0.45~km~s$^{-1}$ for $\Delta t=120$~s to 0.34~km~s$^{-1}$
for $\Delta t=240$~s reaching the lowest value of about 0.23~km~s$^{-1}$ for
$\Delta t=480$~s. Other statistical parameters such as $v_\mathrm{med}$,
$v_{10}$, $v_\mathrm{max}$, $\sigma^2_v$, and $\sigma_{v}$ follow the same
trend.

The initially increasing values can be explained by the time required for a
solar feature to move from one to the next pixel. Three velocity values have to
be considered: (1) the photospheric sound speed of $\approx$8~km~s$^{-1}$, (2)
the maximum photospheric velocity of  $\approx$2~km~s$^{-1}$ measured by LCT
techniques, and (3) the average speed for the proper motion of the granulation
of $\approx$0.5~km~s$^{-1}$. For $\Delta t=15$~s the average displacement is
around one tenth of a pixel, whereas the numerical accuracy for a single
measurement is only one fifth of a pixel, because the maximum of the
cross-correlation function can only be determined with this precision. Thus, in
this case a solar feature has not sufficient time to move, which results in
underestimating its velocity.

In case of $\Delta t=$~60--90~s, the horizontal displacement is sufficiently
large so that a feature could have moved to one of the neighboring pixels. The
speed in individual LCT maps is now well within the range, where numerical
accuracy issues are negligible. Starting at $\Delta t=120$~s the mean velocity
becomes lower, while there is sufficient time for a solar feature to move quite
some distance, the feature might have evolved too much, so that the LCT
algorithm might not any longer trace the same feature. This leads to diminished
horizontal velocities.

Thus, 60--90~s is the good choice for measuring of horizontal flow speeds with
LCT techniques. In this range of the time cadence, the mean values $\bar{v}$ of
the interleaved data sets are essentially the same. Their deviations are much
smaller than the previously discussed systematic trends. In summary, all flow
maps for the database were calculated using $\Delta t=$~60--90~s. If the time
interval $\Delta t$ was shorter, a multiple of the time interval $\Delta
t^{\prime} = n \Delta t$ with $n = 2, 3,$ or 4 was used.

\subsection{Determining the duration of the time averages\label{SEC4.2}}

How many individual LCT maps have to be averaged to yield a reliable flow map?
As previously discussed, the parabola fit to the maximum of the
cross-correlation sets one limitation. However, there are also
method-independent issues to be considered. Solar features evolve over time so
that a global pattern reveals itself only after averaging many individual LCT
maps. We computed the mean horizontal flow speed as a function of the number of
individual LCT maps that were used to arrive at an average flow map. The number
of flow maps corresponds to the time interval $\Delta T$ over which the
individual flow maps were averaged.

Figure~\ref{FIG03} presents this functional dependence for the time cadences
from $\Delta t = 15$~s to 480~s. All curves start with high velocities when only
a few individual LCT maps are averaged. It takes about 20~min before the curves
level out and approach an asymptotic value indicating that these average flow
maps are still dominated by the motions of fine structure contained within the
sampling window. As discusses in the context of the statistical properties of
the flow maps, short time cadences $\Delta t$ tend to underestimate the flow
speed. If flow speeds have be computed for time intervals shorter than 20~min,
in particular for small-scale features, feature tracking methods are more
appropriate than LCT techniques.

% Figure 4
\begin{figure}[t]
\includegraphics[width=0.5\textwidth]{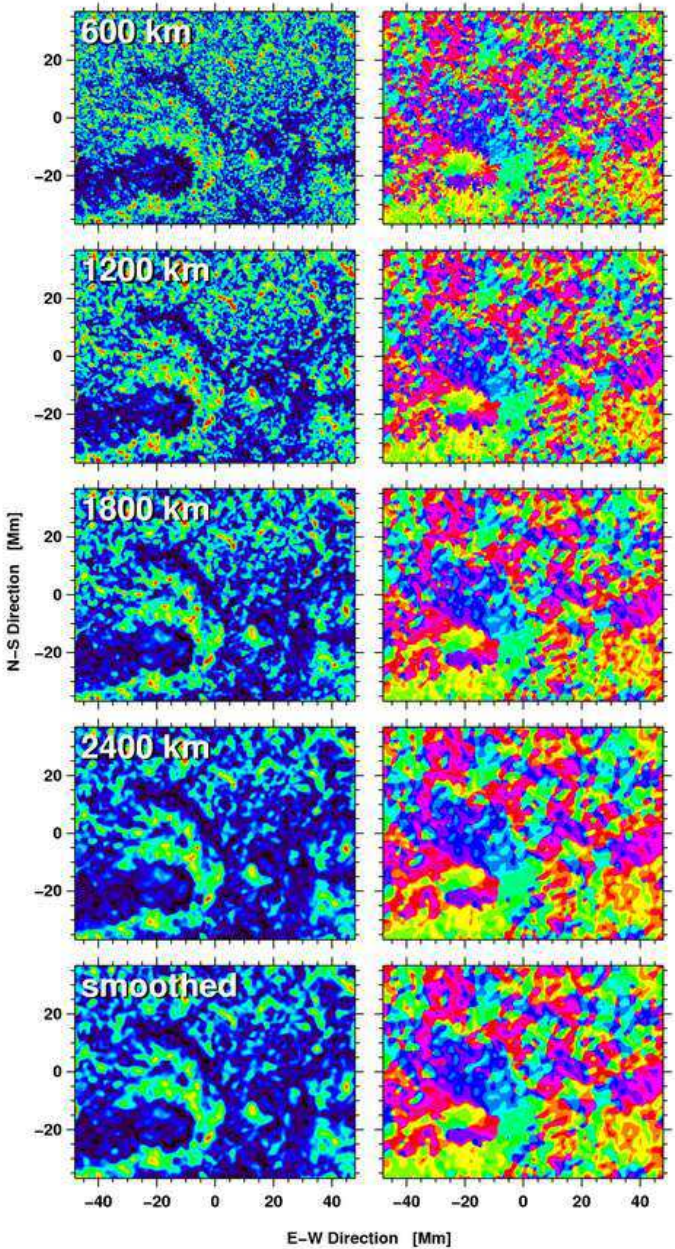}
\caption{Horizontal flow speed (\textit{left}\tsp) and azimuth maps
    (\textit{right}\tsp) measured using Gaussian kernels with $64 \times 64$
    pixels and various FWHM. The FWHM of 7.5, 15, 22.5, and 30 pixels correspond
    to 600, 1200, 1800, and 2400~km on the solar surface, respectively.
    The bottom row depicts smoothed versions of the $\mathrm{FWHM} = 1200$~km
    speed and azimuth maps. The color coding is the same as in
    Fig.~\ref{FIG02}.}
\label{FIG04}
\end{figure}

We find that all curves up to $\Delta t = 90$~s are stacked on top of each other
without crossing the next higher curve at any point. Starting at $\Delta t =
120$~s we find that the curves corresponding to the longer time cadences cross
the other curves after about 20--25~min. This is another indication that solar
features have evolved too much so that LCT fails to properly track their motion.
This behavior provides an explanation for the spread of velocity values found in
literature. In particular, short time-series, as often encountered in
ground-based observations, might be biased towards higher velocities. In
summary, our choice of 60-minute averages for LCT maps is a conservative one
giving solar features sufficient time to reveal the global flow pattern.

\subsection{Selection of the sampling window\label{SEC4.3}}

How do the horizontal proper motions depend on size and FWHM of the Gaussian
kernel used in the LCT algorithm? To answer this question, we calculated
horizontal proper motions using a Gaussian kernel with $64 \times 64$ pixels,
which is equivalent to 5120~km $\times$ 5120~km on the solar surface. This
larger kernel was chosen to encompass successively broader FWHM. We choose four
FWHM of 7.5, 15, 22.5, and 30~pixels corresponding to 600, 1200, 1800, and
2400~km, respectively. The FWHM of 1200~km matches the size of a granule.
Individual LCT maps are produced from image pairs separated by 60~s in time. We
computed the statistical parameters relating to granulation for the entire
sequence and the interleaved data sets. All statistical
parameters are decreasing with increasing FWHM. For $\mathrm{FWHM} = 600$~km
the mean velocity is $\bar{v}=0.48$~km~s$^{-1}$, which decreases to
$\bar{v}=0.47$~km~s$^{-1}$ for $\mathrm{FWHM} = 1200$~km, $\bar{v} =
0.41$~km~s$^{-1}$ for $\mathrm{FWHM} = 1800$~km, and further to $\bar{v} =
0.37$~km~s$^{-1}$ for $\mathrm{FWHM} = 2400$~km. Small-scale features, which
exhibited the highest proper motions, are lumped together with regions of low
flow speeds when the FWHM increases. This effect is displayed in
Fig.~\ref{FIG04}, which shows the speed and azimuth maps for the four FWHM. As
before, for a given FWHM all statistical parameters agree with each other for up
to three significant digits, i.e., there is no indication that the algorithm's
numerical accuracy depends on the FWHM of the kernel used in LCT.

In the case of $\mathrm{FWHM} = 1200$~km, the statistical parameters are very
close to the ones calculated for the same FWHM but with a kernel of $32 \times
32$ pixels. This is not surprising, since the Gaussian kernel assigns a much
stronger weight to features in the center of the subimage, i.e., the periphery
in the $64 \times 64$ pixels FOV has only a small influence in determining the
displacement vector for a pair of subimages. This of course only holds true as
long as the wings of the Gaussian do not significantly extend beyond the edges
of the kernel. Convolving the average flow map ($\mathrm{FWHM} = 1200$~km) with
a Gausssian kernel, which had a size of $32 \times 32$ pixels and $\mathrm{FWHM}
= 26.4$ pixels (2112~km), we arrive at a smoothed version (see bottom row in
Fig.~\ref{FIG04}), which is virtually identical to the flow map with
$\mathrm{FWHM} = 2400$~km.

In conclusion, it makes no difference, if one uses a larger FWHM while computing
LCT maps or one smoothes the maps after computation. In both cases, the results
are virtually the same. Since only minor changes in the LCT results were
observed for kernels with $32 \times 32$ pixels as compared to $64 \times 64$
pixels, the smaller kernel was chosen significantly reducing the computing time.
Furthermore, the smallest FWHM produces the most detailed flow maps. However, we
chose a FWHM of 1200~km favoring the spatial scales of granulation, which covers
the largest fraction of the observed area. Additional smoothing can still be
applied in the later data analysis stages to either reduce noise or to track
flows on larger spatial scales. For case studies of sunspots fine structure
smaller FWHM might be more appropriate.

\subsection{Numerical errors in calculating flow maps\label{SEC4.4}}

We computed the pixel-to-pixel rms-error for the magnitude and direction of the
flow velocity using the interleaved data sets for different time cadences
$\Delta t$ and for different FWHM of the sampling window. Six difference maps
(sets $1-2$, $1-3$, $1-4$, $2-3$, $2-4$, and $3-4$) can be computed from the
four interleaved data sets, thus for each pixel we can derive the errors, which
are primarily due to numerical errors inherent to the LCT algorithm. In case of
different time cadences, the rms-error in velocity is 15--90~m~s$^{-1}$ from
shortest to longest cadence. The corresponding rms-error in direction is
$5^{\circ}$--$30^{\circ}$. However, for cadences in the range of 60--90~s, which
is the range used to create the database, the rms-error of the velocity is
typically in the range from 35--70~m~s$^{-1}$, while the values for the
direction vary by as much as $10^{\circ}$--$15^{\circ}$. The largest variations
in direction are observed near the boundaries of patches showing coherent flows.
Even after correcting the $2\pi$ ambiguity in the difference maps, we find high
values at these locations. As a side note, the rms-error in direction justifies
our choice of a color wheel with only twelve segments in the display of the
azimuth maps. One segment covers $30^{\circ}$ so that pixel-to-pixel variations
of about $\pm 15^{\circ}$ are suppressed. Otherwise, the azimuth maps would
appear too crowded distracting from the overall flow pattern.

% Figure 5
\begin{figure*}[t]
\centerline{\includegraphics[width=0.88\textwidth]{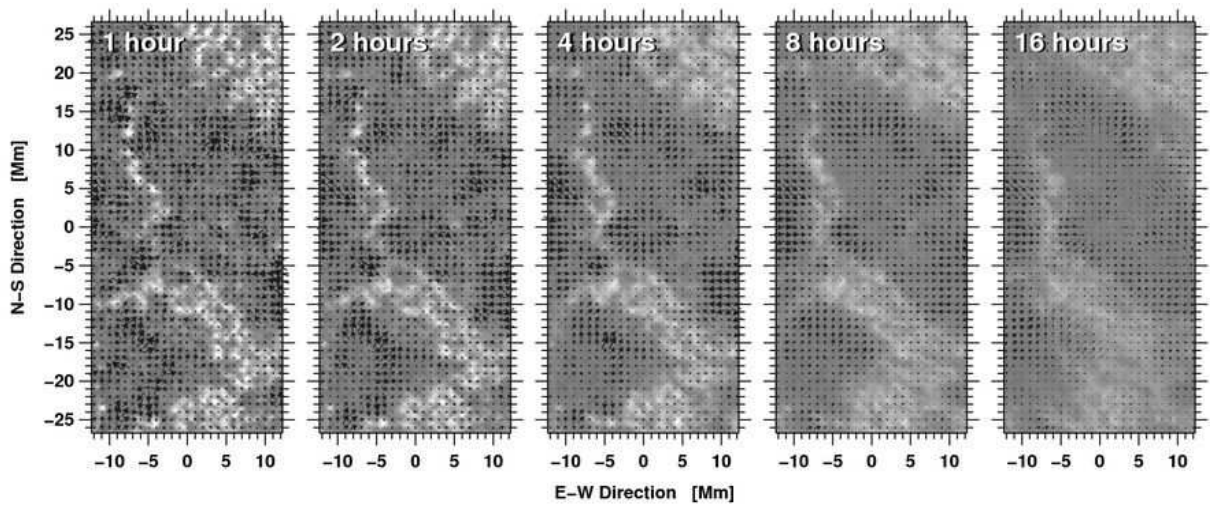}}
\caption{Time-averaged G-band images with horizontal flow vectors computed
    for $\Delta T=$ 1, 2, 4, 8, and 16~h. Arrows with a length corresponding
    to the grid spacing indicate velocities of 1~km~s$^{-1}$. The background
    images are displayed in the intensity range $0.5 \leq I / I_0 \leq 1.5$.}
\label{FIG05}
\end{figure*}

In the case of different FWHM of the sampling window, rms-errors for speed and
direction decrease with increasing FWHM. The velocity error is 35~m~s$^{-1}$ and
the error in direction is about $10^{\circ}$ for a FWHM of 1200~km. These errors
decreases to 15~m~s$^{-1}$ and $5^{\circ}$ for a FWHM = 1800~km and further to
10~m~s$^{-1}$ and $4^{\circ}$ for a FWHM = 2400~km. Here, the decreasing
rms-errors can be attributed to the smoothing effect of a wider sampling window,
i.e., more pixels are used with higher weights in the cross-correlation.
Furthermore, the rms-error in magnitude and direction for the time cadence
$\Delta t = 60$~s is nearly the same regardless of the size of the Gaussian
kernel ($32 \times 32$ pixels vs.\ $64 \times 64$ pixels.

Finally, we calculated Pearson's correlation coefficient between LCT maps of the
interleaved data sets. Pearson's correlation coefficient indicates the degree of
a linear relationship between two variables. A positive value of unity indicates
that the data sets are identical disregarding a linear scaling factor. The
linear correlation coefficient for different time cadences $\Delta t$ decreases
from 0.99 to 0.93 starting at the shortest and ending at the longest cadence.
The high degree of correlation indicates that all essential features of the flow
field are capture by the LCT algorithm. The monotonic decrease, however,
indicates that numerical errors increase when the cadences $\Delta t$ become too
large.

\subsection{Long-lived features in flow maps\label{SEC4.5}}

The one-hour time interval over which the LCT maps are averaged is not
sufficient to identify features, which need longer to evolve such as meso- and
supergranulation. Therefore, to clearly identify the boundaries of these
large-scale convective cells and to visualize the effects of longer time
averages, we averaged LCT maps over $\Delta T =$ 1, 2, 4, 8, and 16~h utilizing
the long-duration data set with 960 images and a time cadence $\Delta t = 60$~s.

As before, we calculated the statistical parameters for the entire sequence and
the interleaved data sets for granulation. All velocity values
decrease with increasing time intervals over which, the individual LCT maps are
averaged. For $\Delta T = 1$~h the mean velocity is $\bar{v} =
0.44$~km~s$^{-1}$, which decreases to $\bar{v} = 0.38$~km~s$^{-1}$ for $\Delta T
= 2$~h, $\bar{v} = 0.34$~km~s$^{-1}$ for $\Delta T = 4$~h, $\bar{v} =
0.30$~km~s$^{-1}$ for $\Delta T = 8$~h, and further to $\bar{v} =
0.23$~km~s$^{-1}$ for $\Delta T = 16$~h. The mean speed approaches a value for
the global flow field with increasing $\Delta T$. However, the value for $\Delta
T = 1$~h is only slightly smaller than previously computed for the high-cadence
sequence. Such small deviations ($<0.05$~km~s$^{-1}$) reflect only minute
differences between the scenes on the solar surface.

% Table 1
%\input{tab01.tex}
\begin{table*}[t]
\begin{center}
\caption{Parameters describing the horizontal proper motions of
         granulation and G-band bright points calculated over 
         $\Delta T = 1, 2, 4, 8, \mathrm{and}$ 16~h.}\medskip
\small
\label{TAB01}
\begin{tabular}{rccccccccccc}
\hline\hline
 & \multicolumn{5}{c}{Granulation} & \rule{5mm}{0mm} &
\multicolumn{5}{c}{Bright points}\rule[-3mm]{0mm}{8mm}\\
\cline{2-6} \cline{8-12}
    $\Delta T$ & $\bar{v}$ & $v_\mathrm{med}$ & $v_{10}$ & $v_\mathrm{max}$ & $\sigma_{v}$
    & & $\bar{v}$ & $v_\mathrm{med}$ & $v_{10}$ & $v_\mathrm{max}$ & $\sigma_{v}$ \rule[-2mm]{0mm}{6mm}\\
    \mbox[h] & [km~s$^{-1}$] & [km~s$^{-1}$] & [km~s$^{-1}$] & [km$^2$~s$^{-2}$] & [km~s$^{-1}$]
    & & [km~s$^{-1}$] & [km~s$^{-1}$] & [km~s$^{-1}$] & [km$^2$~s$^{-2}$] & [km~s$^{-1}$] \rule[-2mm]{0mm}{3mm}\\
\hline
1    & $0.43$ & $0.39$ & $0.78$ & $1.86$ & $0.24$& & $0.22$ & $0.22$ & $0.37$ & $0.70$ & $0.10$ \rule{0mm}{4mm}\\
2    & $0.39$ & $0.36$ & $0.69$ & $1.42$ & $0.21$& & $0.22$ & $0.22$ & $0.35$ & $0.55$ & $0.10$ \rule{0mm}{4mm}\\
4    & $0.34$ & $0.31$ & $0.60$ & $1.25$ & $0.19$& & $0.17$ & $0.17$ & $0.27$ & $0.52$ & $0.08$ \rule{0mm}{4mm}\\
8    & $0.30$ & $0.29$ & $0.50$ & $0.87$ & $0.15$& & $0.15$ & $0.15$ & $0.24$ & $0.45$ & $0.07$\rule{0mm}{4mm}\\
16   & $0.23$ & $0.22$ & $0.38$ & $0.74$ & $0.12$& & $0.12$ & $0.12$ & $0.20$ & $0.41$ & $0.06$ \rule{0mm}{4mm}\\
\hline
\end{tabular}
\end{center}
\end{table*}

% Figure 6
\begin{figure*}[t]
\centerline{
\includegraphics[width=0.5\textwidth]{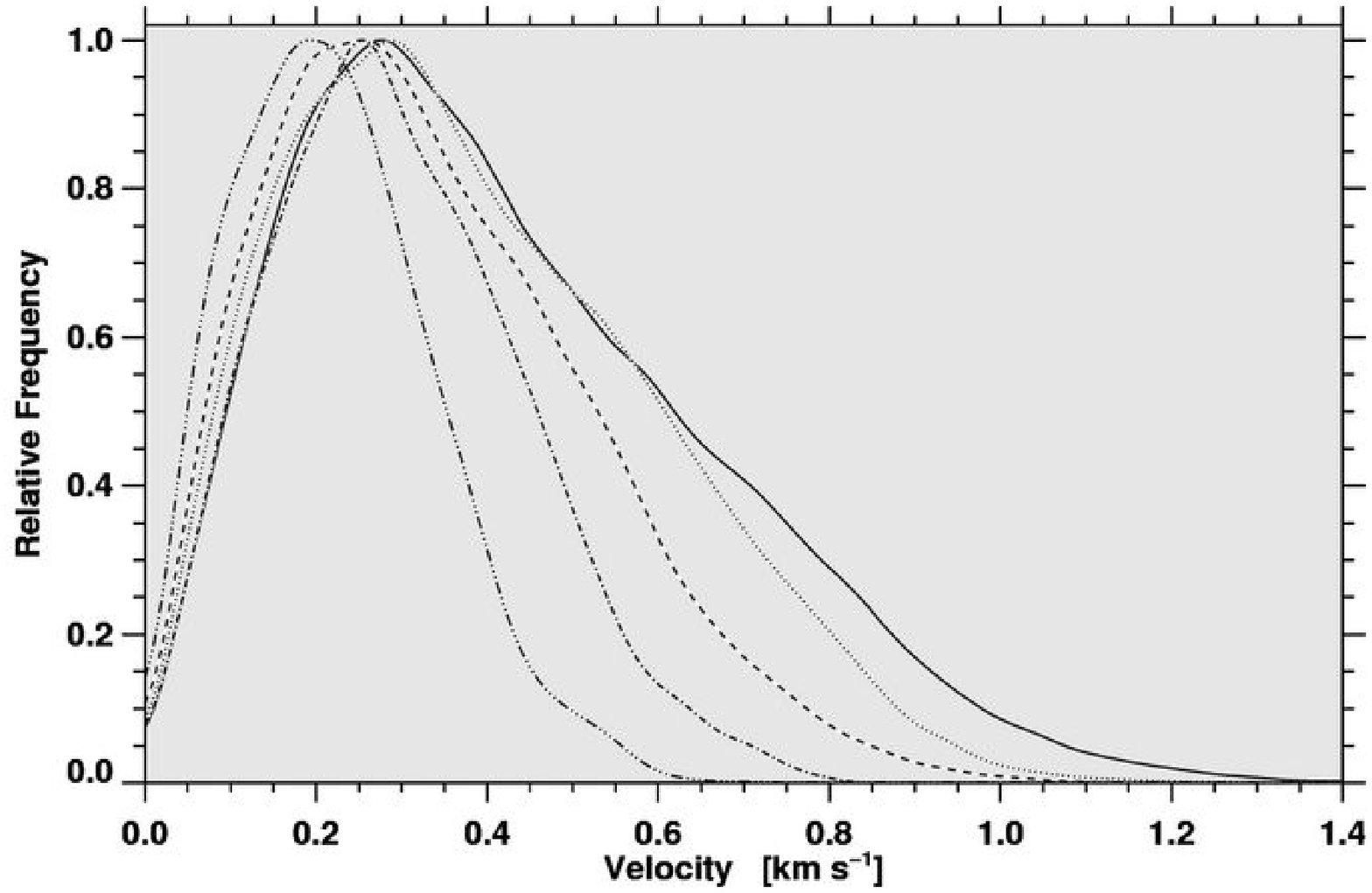}
\includegraphics[width=0.5\textwidth]{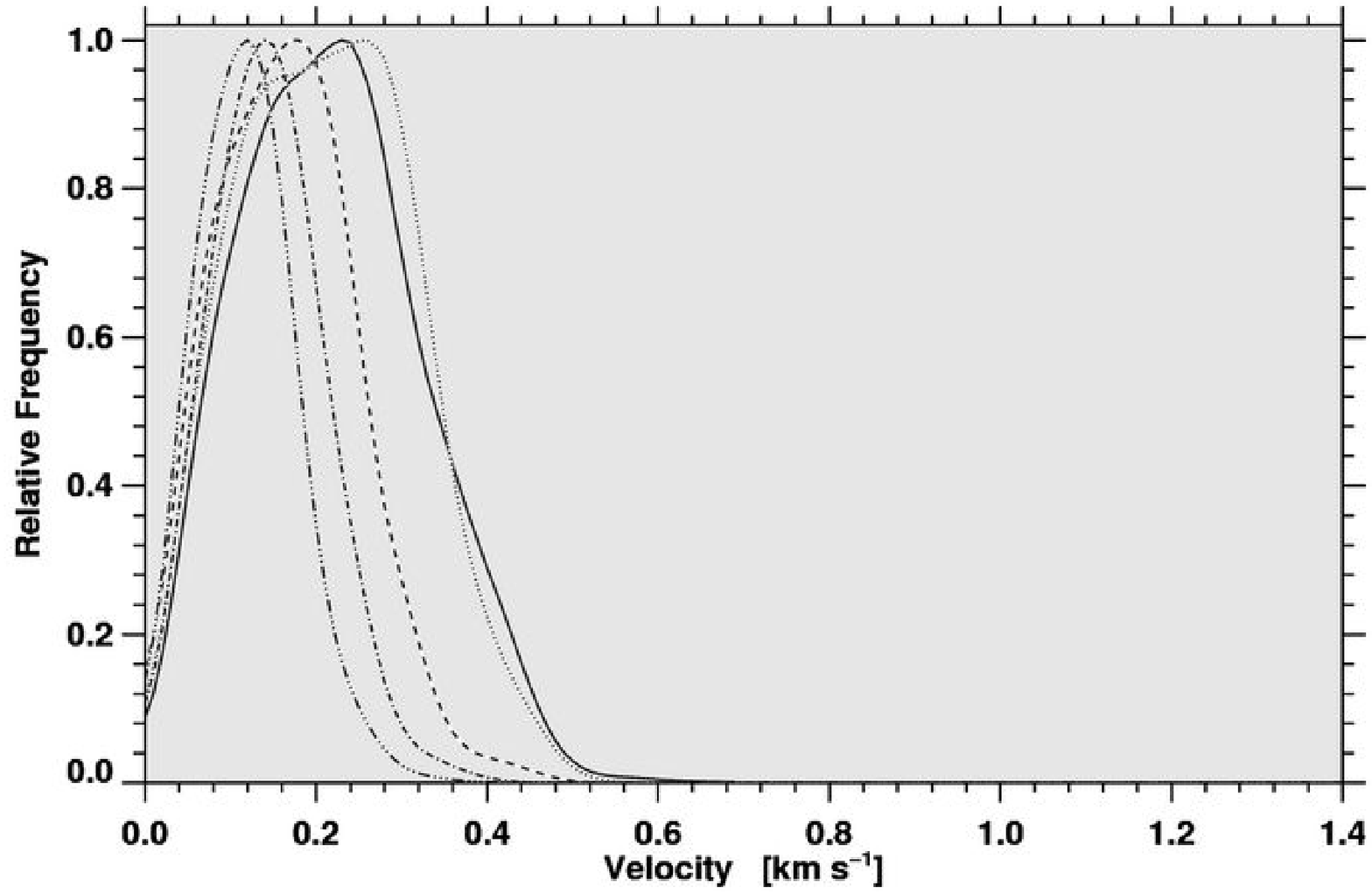}}
\caption{Relative frequency distributions of horizontal proper motions for quiet
    Sun (\textit{left}\tsp) and G-band bright points (\textit{right}\tsp)
    calculated over $\Delta T = 1$~h (solid curve), 2~h (dotted curve), 4~h
    (dashed curve), 8~h (dash-dot-dotted curve), and 16~h (dash-triple dotted
    curve)}
\label{FIG06}
\end{figure*}

Fig.~\ref{FIG05} contains the time-averaged G-band images with superposed arrows
indicating speed and direction of the horizontal proper motions for $\Delta T =
1$, 2, 4, 8, and 16~h. The region shows granulation and G-band bright points to
the west of active region NOAA~11930. Magnetic features dominate the
long-duration time averages, i.e., the flow speed is low where strong magnetic
fields are present in the chromospheric network and conversely the speed is high
where G-band bright points outline the boundaries of large-scale convective
cells. The high speeds associated with the local convective pattern of granules
have diminished for the long-duration flow maps and only the converging motion
towards the cell boundaries remains explaining the statistical properties of the
velocity values discussed above. The overall visual impression of the vector
maps is that the arrows are more ordered in the long-duration maps, whereas in
the short-duration maps ($\Delta T = 1$ and 2~h) a larger scatter of the flow
vectors is observed on smaller scales. Nonetheless, the imprint of the the meso-
and/or supergranulation is already visible in the short-duration flow maps and
becomes more prominent the longer the time interval over which the LCT maps are
averaged. Strong converging motions can be found in Fig.~\ref{FIG05} near the
vertical alignment of G-band bright points in the north-east corner of the FOV
and at the boundary of the larger supergranular cell in the south-east corner of
the FOV. The supergranule also contains substructure on smaller scales, i.e., a
strong divergence center exactly in the central FOV, which can be clearly seen
after averaging for at least $\Delta T = 2$~h.

We plotted relative frequency distributions of flow fields spanning time
intervals of $\Delta T = $ 1, 2, 4, 8, and 16~h in Fig.~\ref{FIG06} to gain
insight into the statistical properties of the long-duration data sets -- this
time both for granulation and for G-band bright points. The statistical
parameters characterizing these distributions are provided in Tab.~\ref{TAB01}
for reference. The distribution describing granulation for the shortest time
interval ($\Delta T=1$~h) is the broadest and has an extended high velocity
tail. Interestingly, for velocities up to 0.6~km~s$^{-1}$ this distribution is
virtually the same as the distribution for an averaging time, which is twice as
long ($\Delta T=2$~h). The only difference is the high velocity tail. This
indicates that proper motions on small scales still make their presence known,
if individual LCT maps are not averaged for at least two hours. The longer
$\Delta T$, the more are the peaks of these distributions shifted towards lower
velocity values (from 0.43 to 0.23~km~s$^{-1}$) and their standard deviations
are progressively becoming smaller (from 0.24 to 0.12~km~s$^{-1}$). The
progression of the frequency distributions shown in the left panel of
Fig.~\ref{FIG06} supports the conclusion that the essential features of
long-duration LCT maps have been captured for $\Delta T = $ 8--16~h.

The frequency distributions for G-band bright point differ in some aspects from
those for granulation. The high velocity tail is less prominent and all
statistical parameters describing the distributions are reduced by about a
factor of two. The two distribution with the shortest time intervals ($\Delta
T=1$ and 2~h) show a hint of a bimodal distribution, and they are skewed towards
higher velocity values. However, this might be an artifact of the adaptive
thresholding algorithm, since the areas covered by G-band bright points are
smaller and well defined in the short-duration intensity maps. Thus, considering
the FWHM = 1200~km of the sampling window, a larger contribution from
granulation is expected, if more isolated G-band bright points are present in
the maps, which are used for threshholding. It should be noted that
Eqn.~\ref{EQN04} was slightly modified to accommodate the longer time intervals,
which leads to a fuzzier appearance of the area covered by G-band bright points
and results in a diminished contrast of the G-band bright points. Since no
contemporary magnetograms with comparable spatial resolution were available, we
cannot comment on the influence of flux emergence or dispersal during the
observed time interval. However, active region NOAA~10930 showed pronounced
activity. In particular, the penumbra of the small sunspot just to the east of
the FOV decayed and resulted in continuous flaring in the active region.

Even though long-duration time averages are an important tool when studying
large-scale convective patterns or the persistent motions in an active region,
the scarcity of such data sets argues against their use for comprehensive and
comparative studies. Since most of the characteristics of flow fields are
already captured in one-hour averages, we opted for $\Delta T = 1$~h.
Furthermore, selecting $\Delta T=1$~h allows us to study changes in
long-duration time-series by computing averaged flow maps every 30~min. Another
consideration is that one hour is more then ten times the typical lifetime of
granules so that the proper motions of individual granules should be negligible
and global motion patterns will reveal themselves.

% Table 2
%\input{tab02.tex}
\begin{table*}[t]
\begin{center}
\caption{Statistical parameters describing the frequency distributions of the
    horizontal proper motions of various solar features.}\medskip
\small
\label{TAB02}
\begin{tabular}{lccccccccc}
\hline\hline
     &
    $\bar{v}$ &
    $v_\mathrm{med}$ &
    $v_{10}$ &
    $v_\mathrm{max}$ &
    $\sigma^2_v$ &
    $\sigma_{v}$ &
    $\gamma_{1,v}$ &
    $\gamma_{2,v}$\rule[-2mm]{0mm}{6mm}\\
Feature & [km~s$^{-1}$] & [km~s$^{-1}$] & [km~s$^{-1}$] & [km~s$^{-1}$] & [km$^2$~s$^{-2}$] & [km~s$^{-1}$] & & \rule[-2mm]{0mm}{3mm}\\
\hline
All             & $0.44$ & $0.40$ & $0.83$ & $1.95$ & $0.07$ & $0.27$ & $0.76$ & \phn $0.21$\rule{0mm}{4mm}\\
Granulation     & $0.47$ & $0.43$ & $0.85$ & $1.95$ & $0.07$ & $0.27$ & $0.67$ & \phn $0.08$\rule{0mm}{4mm}\\
Penumbra        & $0.30$ & $0.24$ & $0.62$ & $1.43$ & $0.05$ & $0.23$ & $1.46$ & \phn $2.07$\rule{0mm}{4mm}\\
Umbra           & $0.23$ & $0.19$ & $0.40$ & $1.92$ & $0.04$ & $0.20$ & $3.66$ & $20.44$\rule{0mm}{4mm}\\
Bright points   & $0.23$ & $0.20$ & $0.43$ & $1.21$ & $0.03$ & $0.15$ & $1.43$ & \phn $2.65$\rule{0mm}{4mm}\\

\hline
\end{tabular}
\end{center}
\end{table*}

\subsection{Frequency distributions for different solar features\label{SEC4.6}}

\begin{figure}[t]
\includegraphics[width=0.5\textwidth]{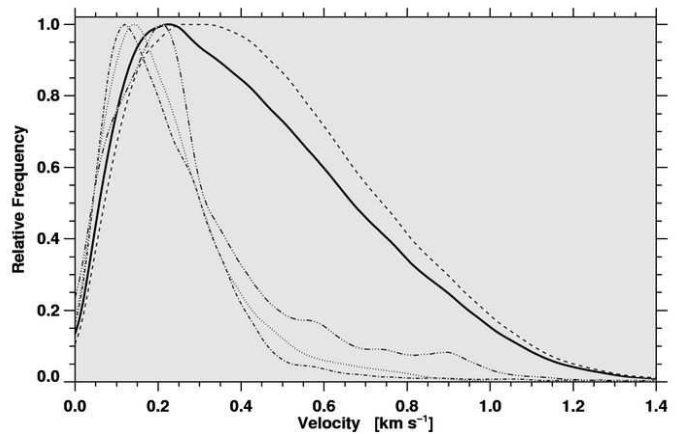}
\caption{Relative frequency distributions of horizontal proper motions computed
     for all solar features (solid curve), G-band bright points (dotted curve),
     granulation (dashed curve), umbra (dash-dotted curve) and penumbra
     (dash-dot-dotted curve).}
\label{FIG07}
\end{figure}

The simplest approach to describe flow fields would be to compute the overall
frequency distributions for a particular FOV. However, this simplistic approach
is insufficient to recover the underlying physics of plasma motions in the
presence (or absence) of strong magnetic field. We used the adaptive
thresholding algorithm (Eqn.~\ref{EQN04}) described in Sect.~\ref{SEC3.4} to
compute frequency distributions for granulation, penumbrae, umbrae/pores, G-band
bright points, and the entire FOV regardless of the features contained in this
region. For this case study, we used the high-cadence sequence (see
Fig.~\ref{FIG02}b for the thresholded image) with $\Delta t = 60$~s, $\Delta T =
1$~h, and a FWHM = 1200~km. The respective distributions are shown in
Fig.~\ref{FIG07} and the corresponding statistical parameters are summarized in
Tab.~\ref{TAB02}. Similar plots and tables will be included in the database
subsuming the more than 200 data sets, which have been analyzed as part of this
study. We provide Fig.~\ref{FIG07} and Tab.~\ref{TAB02} to facilitate comparison
with other case studies. However, these data are not representative (in the
sense of a mean value) for all data sets contained in the database.

A barely detectable shoulder in the frequency distribution for the entire FOV
and the extended high velocity tail hint already that this distribution contains
contributions from various solar features. Its mean velocity $\bar{v}
=0.44$~km~s$^{-1}$ is slightly smaller than the corresponding value for
granulation $\bar{v} =0.47$~km~s$^{-1}$, which dominates the FOV. The
distribution for granulation is broader and a low value of kurtosis
($\gamma_{2,v}=0.08$) leads to a flatter peak, where any indication of a
shoulder is absent. There is a noticeable difference in the distributions of
strong magnetic elements and granulation. The distributions for G-band bright
points, umbral and penumbral regions are narrow, have sharp peaks, and are
shifted towards lower velocities. The mean velocity for these regions varies
from $\bar{v}=0.30$~km~s$^{-1}$ for penumbrae to $\bar{v}=0.23$~km~s$^{-1}$ for
umbrae/pores and G-band bright points. Interestingly, the distributions for
umbrae/pores and G-band bright point are virtually identical, while that for
penumbra has significant contributions at velocities above 0.4~km~s$^{-1}$. This
is indicative of the more complex flow fields in penumbrae, where penumbral
grains move preferentially in the radial direction -- inward in the inner
penumbra and outward in the outer penumbra. This also illustrates that some of
the small-scale horizontal proper motions can be capture with the current
implementation of the LCT algorithm.

\subsection{Flow maps for different spatial resolution\label{SEC4.7}}

Finally, we would like to address the question on how the spatial resolution
affects the determination of the horizontal proper motions. We used the
high-spatial resolution sequence of Sect.~\ref{SEC2.3} and treated it exactly
the same as all the other data with the exception that the G-band images were
sampled at 40, 80, 120, 160, and 200~km after correction of geometrical
foreshortening. Multiple of 40~km were chosen to match the \textit{Hinode}
SOT/BFI pixel size of 0.055\arcsec. Obviously the number of pixels in the
sampling window had to be adjusted. However, shape, size, and FWHM = 1200~km of
the sampling window were not changed.

% Figure 7

\begin{figure*}[t]
\centerline{\includegraphics[width=0.88\textwidth]{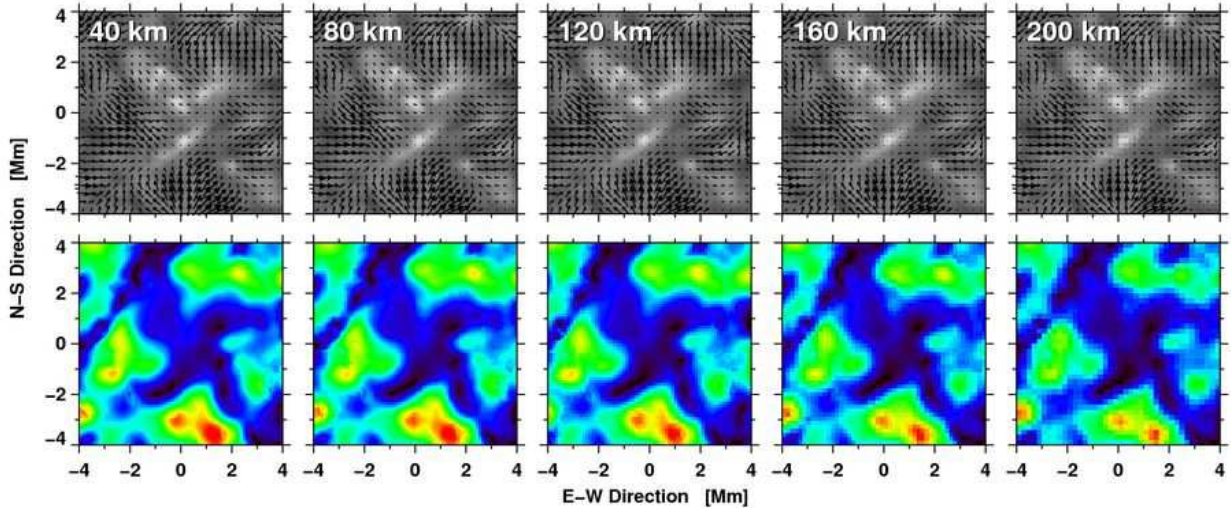}}
\caption{Time averaged G-band images with horizontal flow vectors 
     (\textit{top}\tsp) and flow speed maps (\textit{bottom}\tsp) for
     different spatial resolution. Arrows with the length corresponding
     to the grid spacing indicate velocities of 0.5~km~s$^{-1}$.}
\label{FIG08}
\end{figure*}

The values describing the respective frequency distributions are given in
Tab.~\ref{TAB03}. They were computed for areas with granulation covering the
full FOV. However, to visualize the minute changes in the flow maps, we show in
Fig.~\ref{FIG08} only an area of 8~Mm $\times$ 8~Mm. In the top row of
Fig.~\ref{FIG08} flow vectors are superposed on one-hour average G-band images
with different image scales. This scene on the Sun is dominated by converging
motions towards the \textsf{X}-shaped alignment of G-band bright points. The
grid spacing corresponds to 320~km and the length of the arrows was chosen so
that an arrow with a length of exactly the grid spacing corresponds to
0.5~km~s$^{-1}$. The difference are so minute that they only show up in
difference images of the LCT maps. The bottom row of Fig.~\ref{FIG08} shows the
flow speed for each pixel in the FOV at the same color scale as used in all the
other figures. The overall appearance to the flow field is the same. However,
the low resolution maps look blockier due to the coarser sampling and some of
the fine structure starts to fade out.

% Table 3
%\input{tab03.tex}
\begin{table*}[t]
\begin{center}
\caption{Statistical parameters describing the frequency distributions of the
    horizontal proper motions for various image scales.}\medskip
\small
\label{TAB03}
\begin{tabular}{cccccccccc}
\hline\hline
    \multicolumn{2}{c}{Image scale $\alpha$}  &
    $\bar{v}$ &
    $v_\mathrm{med}$ &
    $v_{10}$ &
    $v_\mathrm{max}$ &
    $\sigma^2_v$ &
    $\sigma_{v}$ &
    $\gamma_{1,v}$ &
    $\gamma_{2,v}$\rule[-2mm]{0mm}{6mm}\\
    
[pixel$^{-1}$] & [km pixel$^{-1}$] & [km~s$^{-1}$] & [km~s$^{-1}$] & [km~s$^{-1}$] &
[km~s$^{-1}$] & [km$^2$~s$^{-2}$] & [km~s$^{-1}$] & & \rule[-2mm]{0mm}{3mm}\\
\hline
0.055\arcsec & \phn 40 & 0.54 & 0.52 & 0.89 & 1.60 & 0.07 & 0.26 & 0.42 & $-0.09$\rule{0mm}{4mm}\\
0.110\arcsec & \phn 80 & 0.53 & 0.52 & 0.88 & 1.62 & 0.07 & 0.26 & 0.42 & $-0.09$\rule{0mm}{4mm}\\
0.165\arcsec &     120 & 0.52 & 0.51 & 0.87 & 1.58 & 0.07 & 0.26 & 0.46 & $-0.04$\rule{0mm}{4mm}\\
0.220\arcsec &     160 & 0.50 & 0.48 & 0.83 & 1.54 & 0.06 & 0.25 & 0.50 & $\phn 0.05$\rule{0mm}{4mm}\\
0.275\arcsec &     200 & 0.47 & 0.45 & 0.79 & 1.57 & 0.06 & 0.24 & 0.55 & $\phn 0.19$\rule{0mm}{4mm}\\
\hline
\end{tabular}
\end{center}
\end{table*}

% Figure 8

In summary, the average velocity diminishes from 0.54~km~s$^{-1}$ at the highest
spatial resolution to 0.47~km~s$^{-1}$ at the lowest resolution. This trend is
the same for all other parameter with the exception of kurtosis and skewness,
which show some (negligible) scatter. Changes of less than 15\% in velocity
cannot explain the broad range of velocity values for horizontal proper motion
reported in literature. Note, however, that \textit{Hinode} data is not
susceptible to the adverse affect of seeing, i.e., ground-based LCT measurements
will be much more affected depending on the spatial resolution. Even though
seeing should not introduce a systematic bias in LCT
\citep[see][]{November1988}, it will still affect the noise in the LCT
measurements.

%###############################################################################
%#
%#    CONCLUSIONS
%#
%###############################################################################

\section{Conclusions\label{SEC5}}

Many case studies exist in the literature describing horizontal proper motions
based on LCT or FT techniques. Even though most of them agree on the morphology
of the observed flows, significant differences are found when quantifying the
flow properties. Besides obvious differences inherent to the techniques, the
choice of parameters such as sampling window, time cadence, and duration can
significantly impact the outcome. Some results of previous studies are provided
in Tab.~\ref{TAB04} for convenience and to ease comparison with the present
investigation.

We have presented the implementation of an LCT algorithm, which was used to
create a database of flow maps derived from time-series of G-band images
observed with \textit{Hinode}/SOT. The parameter study and error analysis will
also be beneficial to other studies using LCT techniques. Even for observations
from the ground our results provide guidance, since LCT techniques are not
biased by seeing \citep[see][]{November1988} so that our error estimates can be
understood as a lower limit.

Justifying the choice of parameters for LCT and FT algorithms is always a
challenging task, which should be driven by the scientific purpose of the study.
In the present study, the emphasis was on creating a database of flow maps,
which can be used in statistical investigations regardless of the type of solar
feature, location on the Sun, or solar activity. In the following list, we
summarize our choice of LCT parameters.

% Table 4
%\input{tab04.tex}
\newcommand{\rb}[1]{\raisebox{4ex}[-4ex]{#1}}
\newcommand{\hb}[1]{\raisebox{6.5ex}[-6.5ex]{#1}}

\begin{table*}[t]
\begin{center}
\caption{Summary of LCT results in previous studies.}\medskip
\small
\label{TAB04}
\begin{tabular}{ccccccccp{70mm}}
\hline\hline
    $\alpha$ &
    FWHM &
    $d_\mathrm{grid}$ &
    $\Delta t$ &
    $\Delta T$ &
    $\bar{v} \pm \sigma_v$ &
    $v_\mathrm{med}$ &
    $v_\mathrm{max}$ &
    \textbf{Remarks}\rule[-2mm]{0mm}{6mm}\\
\mbox{[pixel$^{-1}$]} & & & [s] & [min] & [km~s$^{-1}$] &
[km~s$^{-1}$] & [km~s$^{-1}$] & \rule[-2mm]{0mm}{4mm}\\
\hline

\multicolumn{8}{l}{\citet{November1988} and \citet{November1989}}
    \rule[-2mm]{0mm}{6mm} & continuum images at $\lambda 517.5$~nm, Universal
    Birefringent Filter (UBF), 12~s exposure time, Dunn Solar
    Tel\-e\-scope/Sacramento Peak\\
\rb{0.250\arcsec} & \rb{3.3\arcsec} & \rb{2.0\arcsec} & \rb{67} & \rb{80} &
    \multicolumn{3}{c}{\rb{0.5--1.0 (for source regions)}} \vspace*{-2mm} & \\

\multicolumn{8}{l}{\citet{Brandt1988}} & broad-band (5.4~nm) images at $\lambda
    469.6$~nm, Solar Optical Universal Polarimeter (SOUP), 20~ms exposure time,
    Swedish Solar Vacuum Telescope/Observatorio del Roque de los Muchachos\\
\hb{0.035\arcsec} & \hb{2.4\arcsec} & \hb{0.8\arcsec} & \hb{60} & \hb{79} &
    \hb{0.67} &  & \hb{\phn 1.2} \vspace*{-2mm} & \\

\multicolumn{8}{l}{\citet{Title1989}} & broad-band (100~nm) images at $\lambda
    600$~nm, SOUP, short-exposure images recorded on photographic film,
    Spacelab~2, provides also proper motion measurements for other FWHM down to
    1.0\arcsec\\
\hb{0.161\arcsec}  & \hb{4.0\arcsec} & \hb{2.5\arcsec} & \hb{60} &  \hb{28} &
    \hb{0.37$\pm 0.45$} &  & \hb{$<$1.2} \vspace*{-2mm} & \\

\multicolumn{8}{l}{\citet{Berger1998}} & G-band images $\lambda
    430.5\pm0.6$~nm, SOUP, 20~ms exposure time, SVST, LCT parameter study and
    comparison of proper motions between granulation and network\\
\rb{0.083\arcsec} & \rb{0.83\arcsec} & \rb{0.4\arcsec} & \rb{24} & \rb{70} &
   \rb{$1.10\pm 1.30$} & \rb{0.70} & \rb{$\sim$4.0} \vspace*{-2mm} & \\

\multicolumn{8}{l}{\citet{Shine2000}} & continuum images near the
    Ni\,\textsc{i} $\lambda$676.8~nm line, Michelson Doppler Imager (MDI),
    Solar Heliospheric Observatory (SoHO), long-duration sequence of 45.5~h\\
\rb{0.600\arcsec} & \rb{4.8\arcsec} & \rb{2.4\arcsec} & \rb{60} & \rb{60} &
    \rb{0.49} & \rb{0.47} & \rb{\phn 1.5} \vspace*{-3mm} & \\

\hline
\end{tabular}
\footnotesize
\parbox{170mm}{\vspace*{-2mm}
\begin{itemize}
\item[Note:] $\alpha$: image scale, FWHM: width of the sampling window,
    $d_\mathrm{grid}$: grid spacing on which LCT flow vectors are computed,
    $\Delta t$: image cadence, $\Delta T$: time interval over which individual
    LCT maps are averaged, $\bar{v} \pm \sigma_v$: average flow speed and
    standard deviation, $v_\mathrm{med}$: median value of the frequency
    distribution, and $v_\mathrm{max}$: highest observed flow speed.
\end{itemize}}
\end{center}
\end{table*}

\begin{itemize}
\itemsep0mm
\item[$\square$] The flow maps are based on time-series of G-band images with
    cadences $\Delta t$ between 60~s and 90~s. If the cadence is shorter,
    features with low velocities cannot be accurately tracked, whereas in
    longer cadences features will have evolved too much so that the algorithm
    might not recognize them any longer. Our cadence selection is conservative
    in the sense that we limit our database in favor of better comparability.
    Note that there is a significant number of G-band time-series with cadences
    of about 2~min, which are not included in our database.
\item[$\square$] The evolution of individual features (granules, bright points,
    penumbral grains, umbral dots, etc.) dominate flow fields on short time
    scales. Therefore, averaging over time scales significantly longer than the
    lifetime of the aforementioned features is necessary to yield the global
    flow field. Our choice of $\Delta T = 1$~h over which the flow maps are
    averaged ensures that the global flow fields has emerged from the motions of
    individual features and that sufficient flow maps were averaged to reduce
    the numerical rms-errors for magnitude and direction of the flow vectors to
    reasonable values of 35--70~m~s$^{-1}$ and 10--15$^{\circ}$,
    respectively. Note that $\Delta T = 1$~h is not an appropriate choice for
    studies focusing on meso- and supergranulation because the associated flow
    pattern is still very noisy and could be more easily perceived in longer
    time averages. However, long-duration
    time-series are rare to facilitate such studies. Whenever, time-series with
    longer durations were available, we computed one-hour flow maps with an
    overlap of 30~min so that the temporal evolution of the flow field can be
    monitored, which is of particular interest for the investigation of
    explosive events such as flares, filament eruptions, and coronal mass
    ejections.
\item[$\square$] In principle, the spatial resolution of \textit{Hinode}/SOT
    would allow to track features, which are smaller than one second of arc. Our
    choice of a Gaussian sampling window with 32 $\times$ 32 pixels and a FWHM
    of 1200~km was again motivated by establishing a database of flow maps for
    statistical studies. Therefore, we used a FWHM, which corresponds
    approximately to the size of a granule, which is one of the `largest'
    elements of solar fine structures. Tracking flows on larger spatial scales
    can still be accomplished by smoothing the flow maps after the fact.
\end{itemize}

In forthcoming studies, we will use the database of flow maps to study the
statistical properties of pores, the motions in sunspot penumbrae, and their
relation to the flow pattern observed in the moat of sunspots. Several years of
G-band time-series and more than 1000 individual flow maps facilitate the study
of such flows during the life cycle of solar features and environment,
i.e., as a function of solar activity or the complexity of the
surrounding magnetic field. Once thoroughly tested, the value-added
\textit{Hinode}/SOT data will be made available as a small project within the
scope of GAVO.

%###############################################################################
%#
%#    ACKNOWLEDGMENTS
%#
%###############################################################################

\begin{acknowledgements}
\noindent \textit{Hinode} is a Japanese mission developed and launched by
ISAS/JAXA, collaborating with NAOJ as a domestic partner, NASA and STFC (UK) as
international partners. Scientific operation of the \textit{Hinode} mission is
conducted by the \textit{Hinode} science team organized at ISAS/JAXA. This team
mainly consists of scientists from institutes in the partner countries. Support
for the post-launch operation is provided by JAXA and NAOJ (Japan), STFC (UK),
NASA, ESA, and NSC (Norway). We would like to thank Drs. K.~G.\ Puschmann and
N.\ Deng for carefully reading the manuscript and for providing comments
significantly enhancing the contents of this article. MV expresses her gratitude
for the generous financial support by the German Academic Exchange Service
(DAAD) in the form of a PhD scholarship.
\end{acknowledgements}

%###############################################################################
%#
%#    BIBLIOGRAPHY
%#
%###############################################################################

%\bibliographystyle{/home/meetu/LaTeX/aa}
%\bibliography{/home/meetu/LaTeX/aa-jour,/home/meetu/LaTeX/meetu}
%\input{ref.tex}
%\clearpage

%###############################################################################
%#
%#    APPENDIX
%#
%###############################################################################

%\onecolumn
%\appendix

%\section{Tables}

% Tabe 1 Appendix
%\input{tab01A.tex}

% Tabe 2 Appendix
%\input{tab02A.tex}

% Tabe 3 Appendix
%\input{tab03A.tex}

\end{document}